\documentclass[aps,superscriptaddress,nofootinbib,showpacs,eqsecnum,preprint,tightenlines]{revtex4}
\usepackage{hyperref}
\usepackage{epsfig,rotating}
\usepackage{amsmath,amssymb}
\usepackage{dsfont}

\newcommand{\be}{\begin{equation}}
\newcommand{\ee}{\end{equation}}
\newcommand{\bea}{\begin{eqnarray}}
\newcommand{\eea}{\end{eqnarray}}

\newcommand{\vx}{\vec{x}}

\newcommand{\vq}{\vec{q}}

\newcommand{\vk}{\vec{k}}

\begin{document}

\title{Spontaneous symmetry breaking in inflationary cosmology:\\ on the fate of Goldstone Bosons.}
\author{Daniel Boyanovsky}
\email{boyan@pitt.edu} \affiliation{Department of Physics and
Astronomy, University of Pittsburgh, Pittsburgh, PA 15260, USA}

\date{\today}

\begin{abstract}
We argue that in an inflationary cosmology   a consequence of the lack of   time translational invariance is that spontaneous breaking of a continuous symmetry and Goldstone's theorem \emph{do not} imply the existence of \emph{massless} Goldstone modes. We study spontaneous symmetry breaking in an $ {O}(2)$ model, and implications for $ {O}(N)$ in de Sitter space time. The Goldstone mode acquires a radiatively generated mass as a consequence of   infrared divergences, and the continuous symmetry is spontaneously broken for any finite $N$, however there is a \emph{first order phase transition} as a function of the Hawking temperature $T_H=H/2\pi$. For $O(2)$ the symmetry is spontaneously broken for $T_H < T_c= \lambda^{1/4} v/2.419$ where $\lambda$ is the quartic coupling and $v$ is the tree level vacuum expectation value and the Goldstone mode acquires a radiatively generated mass $\mathcal{M}^2_\pi \propto \lambda^{1/4} H$. The first order nature of the transition is a consequence of the strong infrared behavior of minimally coupled scalar fields in de Sitter space time, the jump in the order parameter at $T_H=T_c$ is $\sigma_{0c} \simeq 0.61\, {H}/{\lambda^{1/4}}$. In the strict $N\rightarrow \infty$ the symmetry cannot be spontaneously broken.  Furthermore, the lack of kinematic thresholds imply that the Goldstone modes \emph{decay} into Goldstone and Higgs modes by emission and absorption of superhorizon quanta.
\end{abstract}

\pacs{98.80.-k,98.80.Cq,11.10.-z}

\maketitle

\section{Introduction}\label{sec:intro}
 In its simplest realization inflationary cosmology can be effectively described as a quasi-deSitter space time.  Early studies\cite{polyakov1,IR1,IR2,allen,folaci,dolgov} revealed that de Sitter space time  features infrared instabilities and profuse particle production in interacting field theories. Infrared divergences in loop corrections to correlation functions hinder the reliability of the perturbative expansion\cite{weinberg,seery,branrecent},  led to the suggestion of an infrared instability of the vacuum\cite{polyakov,kroto,akhmedov,higuchi,vidal}, and affect  correlation functions during inflation\cite{weinberg,giddins,seery,bran,mazumdar,leblond2,woodard,marolf} requiring a non-perturbative treatment.

 Back reaction from particle production in a de Sitter background has been argued to provide a  dynamical``screening'' mechanism that leads to relaxation of the cosmological constant\cite{emil,IR3,branmore}, a suggestion that rekindled the interest on infrared effects in de Sitter space time. A body of work established that infrared and secular divergences are manifest in super-Hubble fluctuations during de Sitter (or nearly de Sitter) inflation\cite{petri,enq,riotto,holman}, thus a  consistent  program that provides a resummation of the perturbative expansion is required. Non-perturbative methods of  resummation of the secular divergences   have been implemented in several studies in de Sitter space time\cite{boyan}       suggesting  a dynamical generation of mass\cite{holman}, a result that was originally anticipated in the seminal work of ref.\cite{staroyoko}, and explored and extended in ref.\cite{richard}. More recently  a self-consistent mechanism of  mass generation for scalar fields through infrared fluctuations has been suggested\cite{petri,holman,rigo,garb,arai,serreau,raja,prokossb,boywwds}.

The lack of a global time-like killing vector in de Sitter space time  leads to remarkable physical effects,  as it implies the lack of particle thresholds (a direct consequence of energy-momentum conservation) and the decay of   fields even in their own quanta\cite{boyprem,boyan} with the concomitant particle production, a result that was  confirmed in ref.\cite{moschella,akhmedov} and more recently investigated in ref.\cite{donmor,leblond} for the case of heavy fields.

 For light scalar fields in de Sitter space time with   mass   $M \ll H$, it was shown in refs.\cite{boyan}  that the infrared enhancement of self-energy corrections is manifest as   poles in $\Delta = M^2/3 H^2$ in correlation functions and that the most infrared singular contributions to the self-energy can be isolated systematically in an expansion in $\Delta$ akin to the $\epsilon$ expansion in critical phenomena. A similar expansion was noticed in refs.\cite{holman,leblond,rigo,smit,serreau}.

Whereas infrared effects in de Sitter (or quasi de Sitter) cosmology are typically studied via correlation functions,   recently the issue of the time evolution of the \emph{quantum states} has began to be addressed. In ref.\cite{boyhol} the Wigner-Weisskopf method\cite{ww,boyaww} ubiquitous in quantum optics\cite{qoptics} has been adapted and extended as a non-perturbative quantum field theory method in inflationary cosmology to study the time evolution of quantum states. This method   reveals how quantum states \emph{decay} in time, it has been shown to be  equivalent to the dynamical renormalization group in Minkowski space time\cite{drg,boyhol} and has recently been implemented to study the radiative generation of masses and decay widths of minimally coupled fields during inflation\cite{boywwds}.

Early studies\cite{ford,ratra} suggested that infrared divergences during inflation can prevent spontaneous symmetry breaking, however more recently the issue of spontaneous symmetry breaking during inflation has been revisited in view of the generation of masses by radiative corrections\cite{serreau,prokossb,arai}. In ref.\cite{serreau} the study of an $O(N)$ model in the large N limit reveals that there is no spontaneous symmetry breaking as a consequence of the infrared divergences: if the $O(N)$ symmetry is spontaneously broken there would be \emph{massless} Goldstone bosons which lead to strong infrared divergences, the resolution, as per the results of this reference is that the symmetry is restored by the strong infrared divergences and no symmetry breaking is possible. This result is in qualitative agreement with those of earlier refs.\cite{ford,ratra}. However, a different study of the same model in ref.\cite{prokossb} reaches a different conclusion: that indeed the $O(N)$ symmetry is spontaneously broken but Goldstone bosons acquire a radiatively induced mass. In ref.\cite{arai} a scalar model with $Z_2$ symmetry is studied with the result that radiative corrections tend to restore the symmetry via the non-perturbative generation of mass. Both refs.\cite{prokossb,arai} suggest a discontinuous transition.

 \vspace{2mm}

 \textbf{Motivation, goals and results:}

 Spontaneous symmetry breaking is an important ingredient in the inflationary paradigm, and as such it merits a deeper understanding of whether radiative corrections modify the familiar picture of slow roll inflation. If, as found in ref.\cite{serreau}, symmetry breaking is not possible in some models, these would be ruled out at least in the simple small field scenarios of slow roll, as inflation would not be successfully ended by the inflaton reaching the broken symmetry minimum. Furthermore, if the inflaton is part of a Higgs-type mode of multiplet of fields, the question of whether the fields associated with unbroken generators are massless is very important as these could lead to entropy perturbations whose infrared divergences are more severe than those of adiabatic perturbations\cite{branrecent}.

 \vspace{2mm}

 In this article we study an $O(2)$ scalar field theory in de Sitter space time and extract implications for $O(N)$ with the following \textbf{goals}: i) to revisit at a deeper level the content of Goldstone's theorem in an \emph{expanding cosmology} in absence of manifest time translational invariance. In particular whether spontaneous symmetry breaking of a continuous symmetry does imply the existence of massless Goldstone modes in an inflationary setting. ii) a   study beyond the local mean field approximation of whether a continuous symmetry can be spontaneously broken in de Sitter space time,  iii) how the mechanism of self-consistent non-perturbative mass generation can be compatible with symmetry breaking and Goldstone modes.

 Recently there has been renewed interest in a deeper understanding of Goldstone's theorem and spontaneous symmetry breaking both in relativistic and non-relativistic systems\cite{brauner,nicolis,wata}, thus our study   provides a complementary investigation of symmetry breaking in a \emph{ cosmological setting} wherein the lack of a global time-like Killing vector leads to unexpected  yet very physical consequences.

 \vspace{2mm}

 \textbf{Brief summary of results:}

 \vspace{2mm}

 \begin{itemize}

 \item{We argue that in absence of time translational invariance Goldstone's theorem \emph{does not} imply the existence of massless excitations if a continuous symmetry is spontaneously broken.  We revisit the implementation of Goldstone's theorem in a spontaneously broken $O(2)$ symmetry in Minkowski space time and highlight that the masslessness of Goldstone Bosons is a consequence of a cancellation between \emph{space time local and non-local terms} in the loop expansion and discuss the implications for an $O(N)$ theory in the large N limit. }

  \item{We then study the same model in de Sitter space-time, and emphasize that whereas in Minkowski space-time the conservation of the Noether current associated with the continuous symmetry directly leads to Goldstone's theorem, in an expanding cosmology this current is \emph{covariantly conserved} and the consequences are, therefore, much less stringent. In conformal coordinates a \emph{conserved} Noether current is manifestly obtained, but the lack of time translational invariance renders the content of Goldstone's theorem much less stringent.}

      \item{ We implement a self-consistent non-perturbative approach based on the Wigner-Weisskopf method  described in refs.\cite{boyhol,boywwds} that allows to extract the mass of the single particle excitations and distinctly shows that the space-time local terms cannot be cancelled by non-local self-energy terms in leading order in a $\Delta$ expansion. As a result Goldstone modes  acquire a radiatively generated mass as a consequence of infrared divergences in agreement with the results in refs.\cite{serreau,prokossb}. The lack of a time-like Killing vector entails that there are no kinematic thresholds, and as a consequence Goldstone modes acquire a \emph{width} from processes of absorption and emission of superhorizon quanta of both Goldstone and Higgs-like modes.  }

  \item{ We show that for finite $N$ there is a symmetry breaking first order transition as a function of the Hawking temperature $T_H = H/2\pi$, Goldstone modes acquire a radiatively infrared generated self consistent mass but also a \emph{decay width}, and that the symmetry cannot be spontaneously broken in the strict $N\rightarrow \infty$ limit. We argue that a first order transition is a distinct and expected consequence of infrared effects, because a continuous transition would entail that at the critical point there should be massless excitations which would lead to infrared divergences. Radiative corrections relieve the infrared singularities by generating a mass but at the expense of turning the symmetry breaking transition into   first order. }

 \end{itemize}

\section{\label{sec:mass} Spontaneous symmetry breaking and Goldstone Bosons \\ in Minkowski space-time:}\label{sec:minkowski}

\subsection{General aspects:}\label{subsec:general}

We consider the $O(2)$ linear sigma model as a simple example of a scalar theory with spontaneous symmetry breaking (SSB) and extract consequences for the case of $O(N)$ in the large N limit.

The Lagrangian density for the $O(2)$   sigma model is
\be  \mathcal{L}   =   \frac{1}{2}(\partial_\mu \,\sigma)^2 +\frac{1}{2}(\partial_\mu \,\pi)^2- V(\sigma^2+\pi^2) \label{sigmamodel}  \ee which is invariant under the infinitesimal transformations
\be \pi \rightarrow \pi + \epsilon \sigma ~~;~~ \sigma \rightarrow \sigma - \epsilon \pi \label{trafo}\ee with $\epsilon$ a space-time constant infinitesimal angle. The canonical momenta conjugate to the $\pi,\sigma$ fields are respectively,
\be P_\pi(x) = \dot{\pi}(x)~~;~~P_\sigma(x) = \dot{\sigma}(x) \label{canmom}\ee with the equal time canonical commutation relations
\be \Big[P_\pi(\vec{x},t),\pi(\vec{y},t) \Big] = -i\,\delta^{3}(\vec{x}-\vec{y})~~;~~\Big[P_\sigma(\vec{x},t),\sigma(\vec{y},t) \Big] = -i\,\delta^{3}(\vec{x}-\vec{y}) \,.\label{ccr}\ee
The conserved Noether current associated with the global symmetry (\ref{trafo}) is
\be J^\mu(x) = i \Big(\sigma(x)\,\partial^\mu \pi(x) - \pi(x)\,\partial^\mu \sigma(x)\Big) ~~;~~\partial_\mu J^\mu(x) =0 \label{conscur}\ee with the conserved charge
\be Q = i\,\int d^3 x  \Big(\sigma(\vx,t)\,P_\pi(\vx,t)- \pi(\vx,t)\,P_\sigma(\vx,t) \Big) \,.\label{charge}\ee

Consider the following identity resulting from current conservation (\ref{conscur}),
 \be \int d^3x  \langle 0|\big[\vec{\nabla}\cdot\vec{J}(\vx,t),\pi(\vec{y},t')\big]|0\rangle  = \frac{\partial}{\partial t} \int d^3x  \langle 0|\big[ {J^0}(\vx,t),\pi(\vec{y},t')\big]|0\rangle \label{conmut}\ee

Assuming spatial translational invariance we introduce
\be S(\vk;t,t') =  \int d^3x ~ e^{-i\vk\cdot(\vec{x}-\vec{y})}\,\langle 0|\big[ {J^0}(\vx,t),\pi(\vec{y},t')\big]|0\rangle \label{Skoft} \ee \emph{If} the surface integral on the left hand side of eqn. (\ref{conmut}) vanishes, then it follows that
\be  {\mathrm{lim}}_{k\rightarrow 0} ~~ \frac{\partial}{\partial t} S(\vk;t,t') = 0 \label{lim1}\ee In general this result implies that
\be {\mathrm{lim}}_{k\rightarrow 0}~~   S(\vk;t,t') = \langle 0|\big[ Q(t),\pi(\vec{y},t')\big]|0\rangle  =  \langle 0|\sigma(\vec{y},t')|0\rangle = v(t') \,.\label{res1}\ee namely $Q$ is time independent.  In absence of time translational invariance the results (\ref{lim1},\ref{res1}) are the only statements that can be extracted from the conservation of the current. However if \emph{time tranlational invariance holds} then $S(\vk;t,t') = S(\vk;t-t')$ and introducing the  spectral representation
\be S(\vk,t-t') = \int \frac{d\omega}{2\pi} S(\vk,\omega) ~~e^{-i\omega(t-t')} \label{specrep}\ee   it follows from (\ref{lim1}) that i) $v(t')=v$ in (\ref{res1}) is time independent and ii)
\be {\mathrm{lim}}_{k\rightarrow 0}~~   S(\vk;\omega) = 2\pi \, v\,\delta(\omega)~~;~~ v = \langle 0|\sigma(\vec{0},0)|0\rangle \,, \label{finres}\ee where we have used eqns.(\ref{charge},\ref{ccr}).

When space-time translational invariance is available further information is obtained by writing $S(\vk,\omega)$ in term of a complete set of eigenstates of the momentum and Hamiltonian operators by inserting this complete set of states in the commutators
\be e^{i\,\vec{P}\cdot\vec{x}}\,e^{-iHt}|n\rangle = e^{i\,\vec{p}_n\cdot\vec{x}}\,e^{-iE_nt}|n\rangle  \,,
\label{intstates} \ee from which we obtain
 \bea S(\vk,\omega)    =   2\pi \sum_{n}  &\Bigg\{&\langle 0|J^0(\vec{0},0)|n\rangle\langle n|\pi(\vec{0},0)|0\rangle ~ \delta^3(\vec{p}_n-\vk)\,\delta(E_n -\omega)- \nonumber \\ && \langle 0|\pi(\vec{0},0)|n\rangle\langle n|J^0(\vec{0},0)|0\rangle ~ \delta^3(\vec{p}_n+\vk)\,\delta(E_n +\omega)\Bigg\} \,. \label{sums}\eea Then the result (\ref{finres}) implies an intermediate state with vanishing energy for vanishing momentum. This is the general form of Goldstone's theorem valid even for non-relativistic systems\cite{lange,brauner,nicolis,wata}. The result has a clear interpretation: under the assumption that the current flow out of the integration boundaries vanishes, the total charge is a constant of motion.  \emph{If the theory is manifestly time translational invariant} this automatically implies that $S(\vec{k},t-t')$ in (\ref{Skoft}) does not depend on $t-t'$ by charge conservation, therefore it follows directly that in the limit $k \rightarrow 0$ the spectral density $S(\vk,\omega)$ can \emph{only} have support at $\omega =0$.

 The standard intuitive explanation for gapless long wavelength excitations relies on the fact that  the continuous symmetry entails that the manifold of  minima away from the origin form a continuum of degenerate states. A rigid rotation around the minimum of the potential does not cost any \emph{energy} because of the degeneracy, therefore the energy cost of making a long-wavelength spatial rotation vanishes in the long-wavelength limit precisely because of the degeneracy. Both this argument and the more formal proof (\ref{finres}) rely on the existence of a conserved energy and energy eigenstates, which is not available in the cosmological setting.

 The main reason for going through this textbook derivation of Goldstone's theorem is to highlight that \emph{time translational invariance} is an essential ingredient in the statement that the Goldstone theorem implies a \emph{gapless excitation} if the symmetry is spontaneously broken\footnote{Under the assumption that the current flow out of a boundary vanishes, see discussion in\cite{lange}.}.

 Precisely this point will be at the heart of the discussion of symmetry breaking in inflationary cosmology.

\subsection{Tree level, one-loop and large N:} \label{subsec:treeonelup}

In order to compare the well known results in Minkowski space-time with the case of inflationary cosmology we now study how Goldstone's theorem is implemented at tree   and one-loop levels in the $O(2)$ case,  and in the large N limit in the case of $O(N)$ symmetry,  as this study will highlight the main differences between Minkowski and de Sitter space times.

To be specific, we now consider the $O(2)$ model with potential
\be V(\sigma^2+\pi^2)   =   \frac{\lambda}{8}\Big(\sigma^2 + \pi^2 -\frac{\mu^2}{\lambda} \Big)^2 \label{potential} \ee
Shifting the field
\be \sigma =  \sigma_0 + \chi \label{shiftsig}\ee

the potential (\ref{potential}) becomes
\be V(\chi,\pi) = \frac{M^2_\chi}{2}\,\chi^2 + \frac{M^2_\pi}{2}\,\pi^2 + \frac{\lambda}{2}\sigma_0 J\, \chi + \frac{\lambda}{2}\sigma_0 \, \chi^3 + \frac{\lambda}{2}\sigma_0 \,\pi^2 \chi + \frac{\lambda}{8} \chi^4 +\frac{\lambda}{8} \pi^4 +\frac{\lambda}{4} \chi^2 \pi^2 \label{potafter}\ee
where
\be J= \sigma^2_0 -\frac{\mu^2}{\lambda}~~;~~ M^2_\chi = {\lambda}\,\Big( \sigma^2_0 + \frac{J}{2} \Big)~~;~~ M^2_\pi = \frac{\lambda}{2}\, J \Rightarrow M^2_\chi - M^2_\pi = \lambda \sigma^2_0 \label{massesetc}\ee
The value of $\sigma_0$ is found by requiring that the expectation value of $\chi$ vanishes in the correct vacuum state, thus it departs from the tree level value $\mu^2/\lambda$ by radiative corrections.

\vspace{2mm}

\textbf{Tree level:}

\vspace{2mm}

At tree level $\sigma^2_0 = \mu^2/\lambda~~;~~M^2_\pi =0,M^2_\chi =  \mu^2$, and  the $\pi$ field obeys the equation of motion
\be \ddot{\pi}(\vec{x},t) - \nabla^2 \pi(\vec{x},t) =0 \,.\label{eompi}\ee The $\pi$ field  is quantized in a volume $V$ as usual
\be \pi(\vec{x},t) = \sum_{\vec{k}}\frac{1}{\sqrt{2Vk}}\Big[a_{\vec{k}}\,e^{-i(kt-\vec{k}\cdot{\vec{x}})}+a^\dagger_{\vec{k}}\,e^{i(kt-\vec{k}\cdot{\vec{x}})} \Big]\,.\label{piquant}\ee

The conserved current  (\ref{conscur}) becomes
\be J^\mu = i\, \sigma_0 \,\partial^\mu\pi +  i \,\Big(\chi \,\partial^\mu\pi - \pi \partial^\mu \chi\Big) \label{shiftedcurr}\ee

At tree level only the first term contributes to the spectral density (\ref{sums}), since at this level the $\pi$ field creates a single particle state out of the vacuum, which is the \emph{only} state that contributes to (\ref{sums}). We refer to the first term as $J^\mu_{tl}$ and its conservation is a result of the equation of motion (\ref{eompi}) and $\sigma_0$ being a space-time constant. It is straightforward to find
\be \langle 0|J^0_{tl}(\vec{0},0)|1_{\vec{p}}\rangle\langle 1_{\vec{p}}|\pi(\vec{0},0)|0\rangle  = -\langle 0|\pi(\vec{0},0)|1_{\vec{p}}\rangle\langle 1_{\vec{p}}|J^0_{tl}(\vec{0},0)|0\rangle = \frac{\sigma_0}{2V}\label{treesumrule} \ee where $V$ is the quantization volume. Therefore
\be S(\vec{k},\omega) =  2\pi \sigma_0 \int \frac{d^3p}{(2\pi)^3} ~\frac{1}{2}\, \Big[\delta(p+\omega)\delta^{3}(\vec{p}+\vec{k})+\delta(p-\omega)\delta^{3}(\vec{p}-\vec{k})\Big] \label{sofkw}\ee and
\be {\mathrm{lim}}_{k\rightarrow 0}~ S(\vec{k},\omega) = 2\pi \sigma_0 \, \delta(\omega) \,. \label{sumruletree}\ee

\vspace{2mm}

\begin{figure}[ht!]
\begin{center}
\includegraphics[height=3in,width=4in,keepaspectratio=true]{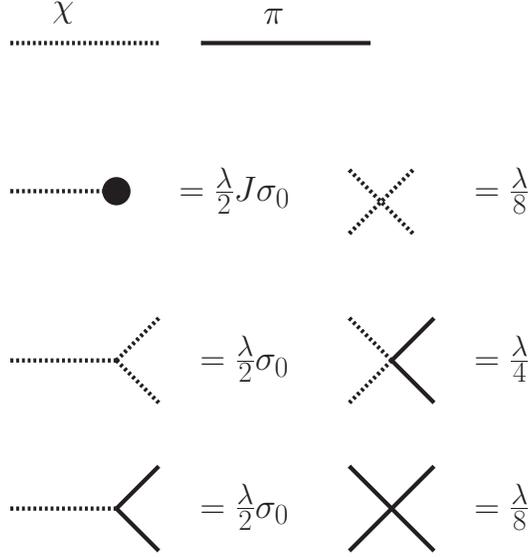}
\caption{Vertices in broken symmetry. The broken line ending in the black dot refers to the \emph{linear} term in $\chi$ in eqn.(\ref{potafter}). }
\label{fig:vertices}
\end{center}
\end{figure}

\textbf{One loop:} We now focus on understanding how the $\pi-$ field remains massless with radiative corrections. We   carry out the loop integrals in four dimensional Euclidean space time, the result is independent of this choice.  The interaction vertices are depicted in fig. (\ref{fig:vertices}).

\begin{figure}[ht!]
\begin{center}
\includegraphics[height=5in,width=4in,keepaspectratio=true]{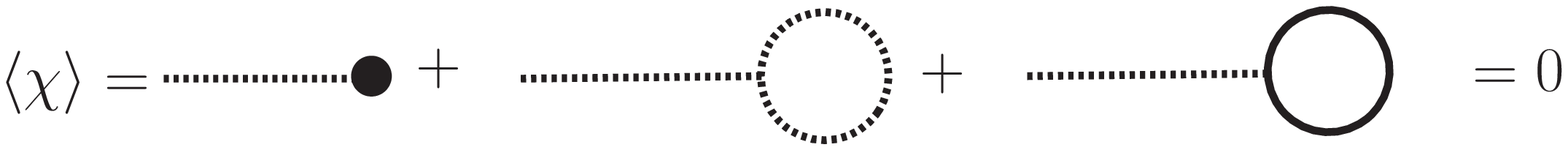}
\caption{Tadpole condition  (\ref{tadcond}). }
\label{fig:expval}
\end{center}
\end{figure}

 The vacuum expectation value $\sigma_0$ is fixed by the requirement that
\be \langle \chi \rangle = 0 \,,\label{tadcond}\ee to which  we refer as the \emph{tadpole condition}, it is depicted in fig.(\ref{fig:expval}). We find
\be \langle \chi \rangle =0 \Rightarrow \frac{\lambda\,\sigma_0}{2\,M^2_\chi}\Big[J + 3 I_\chi + I_\pi \Big] = 0 \label{tadpole}\ee where
\be I_{\chi} = \int \frac{d^4k}{(2\pi)^4}\, \frac{1}{k^2+M^2_\chi} ~~;~~ I_{\pi} = \int \frac{d^4k}{(2\pi)^4}\, \frac{1}{k^2+M^2_{\pi}}\,. \label{tadints}\ee This condition ensures that the matrix element of the interaction Hamiltonian $H_I$ between the vacuum and   single particle states vanishes, namely
\be \langle 1_{\vec{k}}|H_I|0\rangle = 0 \,.\label{nomtxel}\ee

There are two solutions of the tadpole equation
\bea \sigma_0  & = &  0 \,,\label{nossb} \\ J & = & - 3 I_\chi - I_\pi \Rightarrow
\sigma^2_0    =  \frac{\mu^2}{\lambda} - 3 I_\chi - I_\pi \, \neq 0\,, \label{ssbcon}\eea
if available, the second solution (\ref{ssbcon})  leads to spontaneous symmetry breaking.

At finite temperature \be \int \frac{d^4k}{(2\pi)^4}\, \frac{1}{k^2+M^2_{\chi,\pi}} \Rightarrow T \sum_{\omega_n} \frac{d^3k}{(2\pi)^3}\, \frac{1}{\omega^2_n+\vec{k}^2+M^2_{\chi,\pi}}~~;~~ \omega_n = 2\pi\,n\,T \label{finiteT}\ee where $\omega_n$ are the Matsubara frequencies. For $T^2 \gg M^2_{\chi,\pi}$ both integrals are proportional to $T^2$ and the symmetry breaking solution becomes
\be \sigma^2_0 = C\,\Big(T^2_c - T^2\Big) \label{ssbT}\ee with $C$ a positive numerical constant. This well known observation will become relevant below in the discussion of symmetry breaking in de Sitter space time because the (physical) event horizon of de Sitter space-time  $1/H$ determines the Hawking temperature $T_H = H/2\pi$.

\begin{figure}[h!]
\begin{center}
\includegraphics[height=4in,width=4in,keepaspectratio=true]{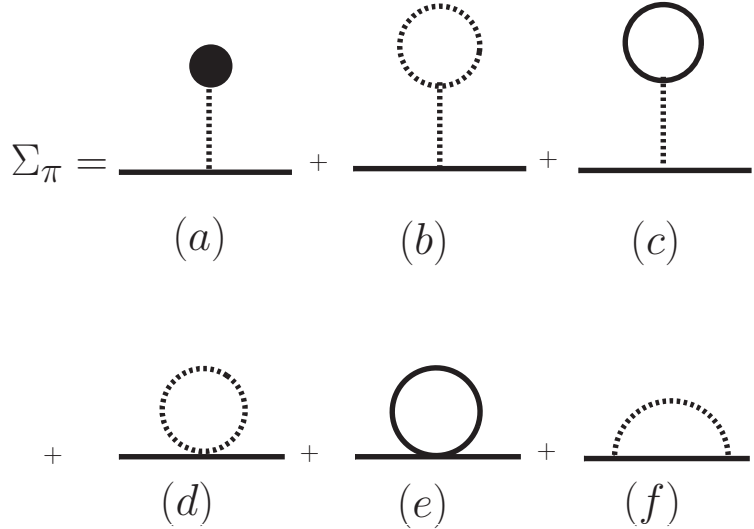}
\caption{One loop diagrams that contribute to the $\pi$ field self-energy $\Sigma_\pi(k)$. }
\label{fig:selfenergy}
\end{center}
\end{figure}

 The $\pi$ propagator becomes
\be G_{\pi}(k) = \frac{1}{k^2+M^2_\pi - \Sigma_\pi(k)} \label{pipropa}\ee where the Feynman diagrams for the self-energy are shown in fig. (\ref{fig:selfenergy}).

The contributions from diagrams (a),(b),(c) yield
\be \Sigma_{\pi,a}(k)+\Sigma_{\pi,b}(k)+\Sigma_{\pi,c}(k)= \frac{\lambda^2 \,\sigma^2_0}{2\, M^2_\chi}  \Big[ J + 3 I_\chi + I_\pi \Big]=0 \label{sigtad}\ee as a consequence of the tadpole condition (\ref{tadpole}). The remaining diagrams yield
\be \Sigma_{\pi,d}(k)+\Sigma_{\pi,e}(k)+\Sigma_{\pi,f}(k)= -\frac{\lambda}{2} \Bigg[I_\chi + 3 I_\pi - 2\lambda \,\sigma^2_0 \int \frac{d^4q}{(2\pi)^4}\frac{1}{(q^2+M^2_\chi)((q+k)^2+M^2_\pi)} \Bigg] \label{siglups}\ee The pole in the $\pi$ propagator   determines the physical mass of the $\pi$ field, we find
\be k^2+ M^2_\pi - \Sigma_\pi(k) = k^2+\frac{\lambda}{2} \Bigg[J+I_\chi + 3 I_\pi - 2\lambda \,\sigma^2_0 \int \frac{d^4q}{(2\pi)^4}\frac{1}{(q^2+M^2_\chi)((q+k)^2+M^2_\pi)} \Bigg] \label{masaphys}\ee where we have used $M^2_\pi$ given by eqn. (\ref{massesetc}).

 If there is spontaneous symmetry breaking, $J = -3I_\chi-I_\pi$  leading to
\be M^2_\pi - \Sigma_\pi(k) =  \lambda\int\frac{d^4q}{(2\pi)^4}\Bigg[\frac{1}{q^2+M^2_{\pi}} - \frac{1}{q^2+M^2_{\chi}}- \frac{\lambda \, \sigma^2_0 }{((q+k)^2+M^2_\pi)(q^2+M^2_{\chi})}\Bigg] \,.\label{massren}\ee Therefore the inverse propagator is given by
\be k^2+ M^2_\pi - \Sigma_\pi(k) = k^2+{\lambda\,\sigma^2_0} \, \int\frac{d^4q}{(2\pi)^4} \frac{1}{q^2+M^2_{\chi}} \Bigg[\frac{1}{q^2+M^2_{\pi}} -   \frac{ 1 }{(q+k)^2+M^2_\pi} \Bigg]  \label{pionmass}\ee where we used eqn. (\ref{massesetc}). Obviously (\ref{massren},\ref{pionmass}) vanish  as $k^2 \rightarrow 0$ (and are proportional to $k^2$ in this limit by Lorentz invariance), therefore the propagator for the Goldstone mode $\pi$ features a pole at $k^2=0$. We emphasize that the vanishing of the mass is a consequence of a precise \emph{cancellation} between the local tadpole terms, fig.(\ref{fig:selfenergy}, (d),(e)) and the non-local (in space-time) contribution fig.(\ref{fig:selfenergy}, (f)) in the $k\rightarrow 0$ limit.

The propagator for $\chi$-the Higgs like mode- is obtained in a similar manner, the Feynman diagrams for the self energy $\Sigma_\chi(k)$ are similar to those for $\Sigma_\pi$ with $\chi$ external lines and the only difference being the combinatoric factors for diagrams (a)-(e), and two exchange diagrams of the (f)-type with intermediate states of two $\chi$ particles and two $\pi$ particles respectively. Again diagrams of the type (a)-(c) are cancelled by the tadpole condition (\ref{tadpole}) and we find
\be k^2+M^2_\chi-\Sigma_\chi(k) = k^2+\frac{\lambda}{2}\Bigg[2  \sigma^2_0 + J + 3 I_\chi +I_\pi - \lambda \sigma^2_0 \,\tilde{I}_\pi(k) -9\,\lambda \sigma^2_0 \,\tilde{I}_\chi(k) \Bigg] \label{chiprop}\ee where
\be \tilde{I}_{\chi,\pi}(k) = \int \frac{d^4q}{(2\pi)^4}\,\frac{1}{\Big(\big(q+k\big)^2+M^2_{\chi,\pi}\Big)^2}\,.\label{tildeI}\ee If the symmetry is spontaneously broken, using the condition (\ref{ssbcon}) we find
\be k^2+M^2_\chi-\Sigma_\chi(k) = k^2 + \lambda \,\sigma^2_0\Big[1- \frac{\lambda}{2}\,\tilde{I}_\pi(k)- \frac{9\,\lambda}{2}\,\tilde{I}_\chi(k) \Big] \label{finchiprop}\ee

\vspace{2mm}

\textbf{Large N limit:}

\vspace{2mm}

If rather than an $O(2)$ symmetry we consider the $O(N)$ case, after symmetry breaking along the $\sigma$ direction the $\vec{\pi}$ fields   belong to an $O(N-1)$ multiplet. In the large N limit the leading term in the tadpole condition $\langle \chi \rangle =0$ (\ref{tadcond}) is given by the last diagram (solid circle) in fig.(\ref{fig:expval}),
\be \langle \chi \rangle =0 \Rightarrow \frac{\lambda\,\sigma_0}{2\,M^2_\chi}\Big[J +  N\, I_\pi \Big] = 0 \label{tadlargeN}\ee where we have neglected terms of $\mathcal{O}(1/N)$ in the large N limit. In this limit the leading contribution to the $\pi$ self-energy is given by fig. (\ref{fig:selfenergy}-(e)),
\be \Sigma_{\pi} = -\frac{\lambda}{2}\, N\,I_\pi \,,\label{sigmapilargeN}\ee where again we neglected terms of $\mathcal{O}(1/N)$. Therefore the inverse $\pi$ propagator in the large N limit is given by
\be k^2+M^2_\pi-\Sigma_{\pi} = k^2 + \mathcal{M}^2_\pi \label{invproplargeN}\ee where
\be \mathcal{M}^2_\pi = \frac{\lambda}{2}\Big[J+N\,I_\pi\Big] \label{ginvlargeNpi}\ee thus   in the large N limit,   the tadpole condition (\ref{tadlargeN}) can be written as
\be \langle \chi \rangle = 0 \Rightarrow \sigma_0 \,\mathcal{M}^2_\pi   = 0 \label{tadpolelarN}\ee therefore if this condition is fulfilled with $\sigma_0 \neq 0$, namely with spontaneous symmetry breaking, automatically the $\pi$ field becomes massless.

\vspace{2mm}

\subsection{Counterterm approach:} \label{subsec:counter}

\vspace{2mm}

An alternative approach that is particularly suited to the study of radiative corrections to masses in the cosmological setting is the familiar method of introducing a mass counterterm in the Lagrangian by writing the mass term in the Lagrangian density as
\be  {M^2_\pi}\pi^2 =   \mathcal{M}^2_\pi \pi^2  +  \delta M^2_\pi \pi^2~~;~~\delta M^2_\pi = M^2_\pi-\mathcal{M}^2_\pi \ee and requesting that the counterterm $\delta M^2$ subtracts the $\pi$ self-energy at zero four momentum
\be -\delta M^2_\pi+\Sigma_\pi(0) =0 \Rightarrow \mathcal{M}^2_\pi = {M}^2_\pi-\Sigma_\pi(0) \label{counter}\ee
and the inverse propagator becomes
\be G^{-1}_\pi(k) = k^2+\mathcal{M}^2_\pi-\Big[\Sigma_\pi(k)- \Sigma_\pi(0)\Big] \label{subprop}\ee in the broken symmetry phase $\mathcal{M}^2_\pi = 0$ from eqns. (\ref{massren},\ref{pionmass}) and the propagator features a pole at zero four momentum.

The main reason to go through this exercise is to highlight the following important points:
\begin{itemize}

\item \textbf{i)} the tadpole type diagrams (a),(b),(c) are cancelled by the tadpole condition (\ref{tadpole}) which is tantamount to the requirement that the interaction Hamiltonian has vanishing matrix element between the vacuum and a single  $\chi$ particle state.

\item     \textbf{ ii)}  at one loop level the vanishing of the $\pi$ mass in the case of spontaneous symmetry breaking is a consequence of the cancellation between the \emph{local} tadpole diagrams (d), (e) and the \emph{non-local} one loop diagram (f) in the $k\rightarrow 0$ limit (the non-locality is in configuration space not in Fourier space). This point will be at the heart of the discussion in inflationary space time below.

\item \textbf{iii)} In the large N limit, only the local tadpole   fig. (\ref{fig:selfenergy}-(e)) contributes to the $\pi$ self-energy and the tadpole condition (\ref{tadpole}), for which a symmetry breaking solution immediately yields a vanishing $\pi$ mass. The tadpole and non-local diagrams fig. (\ref{fig:selfenergy}-(d,f)) are suppressed by a power of $1/N$ in this limit compared to the diagram (\ref{fig:selfenergy}-(e)).

\item \textbf{iv)} The general, non-perturbative proof of the existence of gapless long wavelength excitations as a consequence of the results (\ref{finres},\ref{sums})  manifestly relies on \emph{time translational invariance} and energy eigenstates. In its most general form, without invoking time translational invariance, the result (\ref{res1}) is much less stringent on the long-wavelength spectrum of excitations without an (obvious) statement on the mass spectrum of the theory.  Such a situation, the lack of time translational invariance (global time-like Killing vector) is a hallmark of inflationary     cosmology and it is expected that -unlike in Minkowski space-time- Goldstone modes \emph{may acquire a mass radiatively}.
\end{itemize}

These points are   relevant in the discussion of the fate of Goldstone bosons in de Sitter space-time discussed below.

\section{Goldstone Bosons in de Sitter space-time:}\label{sec:goldcosmo} We consider the $O(2)$ linear sigma model minimally coupled  in a spatially flat de Sitter space time with metric given by
\be ds^2 = dt^2-a^2(t)~ d\vec{x}^2 ~~;~~a(t)=e^{Ht} \label{frw}\ee
defined by the action (the different notation for the fields as compared to the previous section will be explained below)
\be L= \int d^4x \sqrt{|g|}\,\Bigg\{ \frac{1}{2} g^{\mu \nu} \partial_\mu \vec{\Phi}\cdot \partial_\nu \vec{\Phi} - V(\vec{\Phi}\cdot\vec{\Phi}) \Bigg\}~~;~~\vec{\Phi} = (\phi_1,\phi_2) \,.\label{actionds}\ee
were
\be V(\vec{\Phi}\cdot\vec{\Phi}) = \frac{\lambda}{8}\Bigg(\phi^2_1 + \phi^2_2 - \frac{\mu^2}{\lambda}\Bigg)^2\,.
\label{potds}\ee

We follow the method of ref.\cite{coleman} to obtain the conservation law associated with the global $O(2)$ symmetry: consider a space-time dependent infinitesimal transformation that vanishes at the boundary of space-time
\be \phi_1(\vec{x},t) \rightarrow \phi_1(\vec{x},t) - \epsilon(\vec{x},t)\phi_2(\vec{x},t)  ~~;~~ \phi_2(\vec{x},t) \rightarrow \phi_2(\vec{x},t) + \epsilon(\vec{x},t)\phi_1(\vec{x},t) \label{trafods}\ee under which the change in the action is given by
\be \delta L = \int d^4x \sqrt{|g|}\, \partial_\mu \epsilon(\vec{x},t)~J^\mu(\vec{x},t) \label{changeL}\ee where
\be J^\mu(\vec{x},t) = i\,g^{\mu \nu}\Big[\phi_1 \partial_\nu \phi_2 - \phi_2 \partial_\nu \phi_1\Big] \label{current}\ee upon integration by parts assuming a vanishing boundary term,
\be \delta L =  -\int d^4x \sqrt{|g|}\, \epsilon(\vec{x},t)\,J^\mu_{;\,\mu}(\vec{x},t) \label{covacons}\ee from which upon using the variational principle\cite{coleman} we recognize that the current (\ref{current}) is \emph{covariantly conserved}
\be J^\mu_{;\,\mu}(\vec{x},t) = \frac{1}{\sqrt{|g|}}~ \partial_\mu \Big( \sqrt{|g|} \,J^\mu\Big) = \dot{J}^0+3\,H \,J^0-\frac{1}{a^2(t)}\nabla\cdot\Big(\phi_1\,\nabla\,\phi_2-\phi_2\,\nabla\,\phi_1 \Big)=0 \label{covacons2}\ee where the dot stands for $d/d t$. This covariant conservation law can be seen   to follow from the Heisenberg equations of motion for the fields,
\be \ddot{\phi}_a + 3H \dot{\phi}_a - \frac{\nabla^2}{a^2(t)}\phi_a + 2~\Big(\frac{d V(\rho^2)}{d\rho^2}\Big)\,\phi_a = 0 ~~;~~ a= 1,2 ~~;~~ \rho^2 = \phi^2_1 + \phi^2_2 \,. \label{heiseqn}\ee

It is the second term in   (\ref{covacons2}) that prevents a straightforward generalization of the steps leading to Goldstone's theorem as described in the previous section. Fundamentally it is this difference that  is at the heart of the major discrepancies in the corollary of Goldstone's theorem in the expanding cosmology as compared to Minkowski space time.

It is convenient to pass to conformal time
\be \eta = - \frac{e^{-Ht}}{H} ~~;~~ a( \eta ) = -\frac{1}{H\eta} \label{conftime}\ee and to rescale the fields \be \phi_1(\vec{x},t)  = \frac{\sigma(\vec{x},\eta)}{a( \eta)}~~,~~\phi_2(\vec{x},t)  = \frac{\pi(\vec{x},\eta)}{a(\eta)} \label{rescafield}\ee in terms of which the covariant conservation law (\ref{covacons2}) becomes
\be \frac{\partial}{\partial \eta} \mathcal{J}^0(\vec{x},\eta) + \vec{\nabla}\cdot\vec{\mathcal{J}}(\vec{x},\eta) =0 \label{consconf} \ee where
\bea  \mathcal{J}^0(\vec{x},\eta) & = &  i \Big[\sigma \, \pi^{'} -\pi \, \sigma^{'}\Big] \label{joconf}\\
\vec{\mathcal{J}}(\vec{x},\eta) & = &  -i \Big[\sigma \, \vec{\nabla} \pi -\pi \, \vec{\nabla} \sigma \Big] \label{jvecconf}\eea where $'\equiv d/d\eta$.

In terms of the rescaled fields the action becomes (after dropping a total surface term)
\be L=\int d^3x d\eta \Bigg\{\frac{1}{2}\Big[\sigma^{'\,2}-(\nabla \sigma)^2 + \pi^{'\,2}-(\nabla \pi)^2+\frac{a''}{a}(\sigma^2+\pi^2) \Big]-\mathcal{V}\big(\sigma^2+\pi^2;\eta\big) \Bigg\} \label{lagconf} \ee where
\be \mathcal{V}\big(\sigma^2+\pi^2;\eta\big) = \frac{\lambda}{8}\Big(\sigma^2+\pi^2-a^2(\eta) \frac{\mu^2}{\lambda}\Big)^2 \,. \label{potconf}\ee Therefore, although the Noether current (\ref{joconf},\ref{jvecconf}) is conserved and looks similar to that in Minkowski space time, the Hamiltonian is manifestly time dependent, there is no time translational invariance and no energy conservation and no spectral representation is available, all of these are necessary ingredients   for Goldstone's theorem  to guarantee massless excitations.

The Heisenberg equations of motion are
\bea \sigma^{''}-\nabla^2 \sigma + \Big[2 \,\frac{d\mathcal{V}(r^2)}{dr^2}-\frac{a^{''}}{a} \Big]\,\sigma  & =  & 0 \label{sigdseqn}\\ \pi^{''}-\nabla^2 \pi + \Big[2 \,\frac{d\mathcal{V}(r^2)}{dr^2}-\frac{a^{''}}{a} \Big]\,\pi  & =  & 0 \label{pidseqn}
\eea where $r^2 = \pi^2 + \sigma^2$. Using these Heisenberg equations of motion it is straightforward to confirm the conservation law (\ref{consconf}) with (\ref{joconf},\ref{jvecconf}).

Now making an  $\eta$ dependent shift of the field $\sigma$
\be \sigma(\vec{x},\eta) = \sigma_0\,a(\eta)+\chi(\vec{x},\eta) \label{shifteta}\ee the action (\ref{lagconf}) becomes
\bea L & = & \int d^3x d\eta \Bigg\{\frac{1}{2}\Big[\chi^{'\,2}-(\nabla \chi)^2 + \pi^{'\,2}-(\nabla \pi)^2- \frac{1}{\eta^2}\Big(\frac{M^2_\chi }{H^2}-\frac{1}{2}\Big)\,\chi^2-\frac{1}{\eta^2}\Big(\frac{M^2_\pi }{H^2}-\frac{1}{2}\Big)\,\pi^2 \Big]  \nonumber \\ & + &  \frac{\lambda}{2\,\eta^3}\frac{\sigma_0  J}{H^3} \, \chi +\frac{ \lambda}{2\eta}\frac{\sigma_0}{H}  \, \chi^3 + \frac{\lambda}{2\eta}\frac{\sigma_0}{H}  \,\pi^2 \chi - \frac{\lambda}{8} \chi^4 -\frac{\lambda}{8} \pi^4 -\frac{\lambda}{4} \chi^2 \pi^2 \Bigg\} \label{lagrads}
\eea
where $M_{\chi,\pi}, J$ are the same as in the Minkowski space time case given by eqn.  (\ref{massesetc}).
The Heisenberg equations of motion for the spatial Fourier modes
of wavevector $k$  of the fields in the non-interacting ($\lambda=0$)
theory are given by
\bea  \chi''_{\vk}(\eta)+
\Big[k^2-\frac{1}{\eta^2}\Big(\nu^2_\chi -\frac{1}{4} \Big)
\Big]\chi_{\vk}(\eta)  & = &   0   \label{chimodes} \\
\pi''_{\vk}(\eta)+
\Big[k^2-\frac{1}{\eta^2}\Big(\nu^2_\pi -\frac{1}{4} \Big)
\Big]\pi_{\vk}(\eta)  & = &   0   \label{pimodes}
\eea
\noindent where
\be
\nu^2_{\chi,\pi}   =  \frac{9}{4}- \frac{M^2_{\chi,\pi} }{H^2}\,.
   \label{nu}\ee We will focus on the case of ``light'' fields, namely $M^2_{\chi,\pi}\ll H^2$ and
  choose Bunch-Davies vacuum conditions for which the two linearly independent solutions are given by
\bea
g_{\chi,\pi}(k;\eta) & = & \frac{1}{2}\; i^{\nu_{\chi,\pi}+\frac{1}{2}}
\sqrt{-\pi \eta}\,H^{(1)}_{\nu_{\chi,\pi}}(-k\eta)\label{gnu}\\
f_{\chi,\pi}(k;\eta) & = & \frac{1}{2}\; i^{-\nu_{\chi,\pi}-\frac{1}{2}}
\sqrt{-\pi \eta}\,H^{(2)}_{\nu_{\chi,\pi}}(-k\eta)= g^*_{\chi,\pi}(k;\eta) \label{fnu}  \; ,
\eea
 \noindent where $H^{(1,2)}_\nu(z)$ are Hankel functions. Expanding the field operator in this basis in a comoving volume $V$
\bea \chi(\vec{x},\eta)  &  =  & \frac{1}{\sqrt{V}}\sum_{\vec{k}} \Big[a_{\vec{k}}\,g_\chi(k;\eta)\,e^{i\vec{k}\cdot\vec{x}}+ a^\dagger_{\vec{k}}\,\,g^*_\chi(k;\eta)\,e^{-i\vec{k}\cdot\vec{x}}\Big]  \label{quantumfieldchi}\\
\pi(\vec{x},\eta)  &  =  & \frac{1}{\sqrt{V}}\sum_{\vec{k}} \Big[b_{\vec{k}}\,g_\pi(k;\eta)\,e^{i\vec{k}\cdot\vec{x}}+ b^\dagger_{\vec{k}}\,\,g^*_\pi(k;\eta)\,e^{-i\vec{k}\cdot\vec{x}}\Big]  \label{quantumfieldpi}
\eea
The Bunch-Davies vacuum is defined so that \be a_{\vec{k}}|0\rangle = 0 ~~;~~b_k|0\rangle = 0 \,,\label{dsvac}\ee and the Fock  states are obtained   by applying creation operators $a_{\vec{k}}^{\dagger};b^\dagger_{\vec{k}}$ onto the vacuum.

After the shift (\ref{shifteta}), the current (\ref{joconf},\ref{jvecconf}) becomes
\bea \mathcal{J}^0(\vec{x},\eta) & = &  \mathcal{J}^0_{tl}(\vec{x},\eta)+i \Big[\chi \, \pi^{'} -\pi \, \chi^{'}\Big]~~;~~ \mathcal{J}^0_{tl}(\vec{x},\eta) =  i \Big[\sigma_0\,a \, \pi^{'} -\pi \, \sigma_0\,a^{'}\Big] \label{joconfshift}\\
\vec{\mathcal{J}}(\vec{x},\eta) & = &  \vec{\mathcal{J}}_{tl}(\vec{x},\eta) -i \Big[\chi \, \vec{\nabla} \pi -\pi \, \vec{\nabla} \chi \Big]~~;~~\vec{\mathcal{J}}_{tl}(\vec{x},\eta) = -i\sigma_0\,a\,\vec{\nabla} \pi \,.\label{jvecconfshift}\eea The   terms $ \mathcal{J}^0_{tl}(\vec{x},\eta),\vec{\mathcal{J}}_{tl}(\vec{x},\eta) $ on the right hand sides of (\ref{joconfshift},\ref{jvecconfshift}) are the \emph{tree level} contributions to the conserved current as these terms create single particle $\pi$ states out of the vacuum.

The interaction vertices are the same as those for the Minkowski space-time case depicted in fig.(\ref{fig:vertices}) but with the replacements
\be \sigma_0 \rightarrow -\frac{\sigma_0}{H\eta}  ~~;~~J \rightarrow -\frac{J}{H\eta}  \,. \label{repds}\ee

In refs.\cite{boyan,rigo,boywwds} it is found that the tadpole contributions in figs.(\ref{fig:expval},\ref{fig:selfenergy}-(d,e)) are given by
\bea  \langle 0| \chi^2(\vx,\eta)|0 \rangle_{ren} & = &  \frac{1}{8\pi^2 \,\eta^2}~\frac1{\Delta_\chi} ~\big[1+
  \cdots \big]  \label{poletadchi} \\
   \langle 0| \pi^2(\vx,\eta)|0 \rangle_{ren} & = &  \frac{1}{8\pi^2 \,\eta^2}~\frac1{\Delta_\pi} ~\big[1+
  \cdots \big]  \label{poletadpi}\eea where the renormalization regularizes ultraviolet divergences, and
  \be \Delta_\chi  = \frac{M^2_\chi}{3H^2}~~;~~\Delta_\pi  = \frac{M^2_\pi}{3H^2}\,,\label{deltas}\ee   the dots  in eqns. (\ref{poletadchi},\ref{poletadpi}) stand for terms subleading in powers of $\Delta_{\chi,\pi}\ll 1$. In order to maintain a notation consistent with the previous section we introduce
  \be \mathcal{I}_{\chi,\pi} \equiv  \frac{1}{8\pi^2\,\Delta_{\chi,\pi}}  \,.\label{Isdef}\ee
  The tadpole condition now becomes
  \be \langle \chi \rangle = 0 \Rightarrow \frac{\lambda\,a\,\sigma_0}{2\,\eta^2}\Big[\frac{J}{H^2}+3\mathcal{I}_{\chi}+\mathcal{I}_{\pi}\Big] =0 \,. \label{tadpoleds}\ee A symmetry breaking solution corresponds to $\sigma_0\neq 0 ~;~ J/H^2 = -3\mathcal{I}_{\chi}-\mathcal{I}_{\pi} $. At tree level
  \be \sigma^2_0 = \frac{\mu^2}{\lambda} \Rightarrow J=0 \Rightarrow M^2_\pi =0 \,,\label{treelevelssbds} \ee and using that $a^{''}/a = 2/\eta^2$ the tree-level conservation law becomes
  \be \frac{\partial}{\partial \eta}\mathcal{J}^0_{tl} +\vec{\nabla}\cdot\vec{\mathcal{J}}_{tl} = 0 \Rightarrow \sigma_0 a(\eta)\Big[\pi^{''} - \frac{2}{\eta^2}-\nabla^2\pi\Big] =0 \label{tlcl}\ee which is fulfilled by the Heisenberg equation of motion for the $\pi$ field (\ref{pimodes}) with $M_\pi=0$, namely $\nu_\pi = 3/2$.

  It is illuminating to understand how the result (\ref{res1}) is fulfilled at tree level. With  the expansion of the $\pi$ field given by (\ref{quantumfieldpi}) and $\nu_\pi = 3/2$ introduced in $\mathcal{J}^{0}_{tl}(\vec{x},\eta)$ we find
\be S(\vec{k};\eta,\eta') = -2 \,\sigma_0\,a(\eta) \,\mathrm{Im}\Big[g^*_\pi(k;\eta') \Big(g^{'}_\pi(k;\eta)+ \frac{g_\pi(k;\eta)}{\eta} \Big)\Big] \label{skds}\ee and the long wavelength limit is given by
\be \mathrm{lim}_{k \rightarrow 0} \, S(\vec{k};\eta,\eta') = \sigma_0 \,a(\eta')\,. \label{k0lim}\ee

Again, we note that it is precisely the lack of time translational invariance that restricts the content of eqn. (\ref{k0lim}), while this equation is satisfied with $M_\pi =0$ at tree level, there is no constraint on the mass of the single particle excitations from the general result (\ref{res1}). Thus whether the Goldstone fields acquire a mass via radiative corrections now becomes a dynamical question.

There are two roadblocks to understanding radiative corrections to the mass, both stemming from the lack of time translational invariance: i) in general there is no simple manner to resum the series of one particle irreducible diagrams into a Dyson propagator, whose poles reveal the physical mass, ii) there is no Fourier transform in time that when combined with a spatial Fourier transform would allow to glean a dispersion relation for single particle excitations. Obviously these these two problems are related. In refs.\cite{serreau,prokossb,arai} only the local tadpoles were considered, this is a local mean field approximation and the space-time local nature of the tadpole allows to extract a mass. However, while the mean field tadpole is the leading contribution in the large N limit as discussed in the previous section, for finite $N$ the non-local diagram equivalent to fig. (\ref{fig:selfenergy}-(f)) is of the same order, and in Minkowski space time it is \emph{this} diagram that cancels the tadpole (mean field) contribution to the $\pi$ mass. Thus for finite $N$ the question is whether  the non-local self-energy contribution (\ref{fig:selfenergy}-(f)) can cancel the tadpole contributions of fig. (\ref{fig:selfenergy}-(d),(e)) even when these feature very different time dependence and (\ref{fig:selfenergy}-(f)) does not have a time Fourier transform that renders it local in frequency space.

It is at this point where the Wigner-Weisskopf method introduced in refs.\cite{boyhol,boywwds} proves to be particularly useful.

\subsection{Wigner-Weisskopf theory in de Sitter space time:}

In order to make the discussion self-contained, we highlight the main aspects of the Wigner-Weisskopf non-perturbative approach to study the time evolution of quantum states pertinent to the self-consistent description of mass generation discussed in the previous sections. For a more thorough discussion and comparison to results in Minkowski space time the reader is referred to ref.\cite{boyhol,boywwds}.
 Expanding the interaction picture state $|\Psi(\eta)\rangle_I$ in   Fock states $|n\rangle$ obtained as usual by applying the creation operators on to the (bare) vacuum state (here taken to be the Bunch-Davies vacuum) as
  \be |\Psi(\eta)\rangle_I = \sum_n C_n(\eta) |n\rangle \label{expastate}\ee the evolution of the state in the interaction picture given by \cite{boyhol}
  \be i \frac{d}{d\eta}|\Psi(\eta)\rangle_I = H_I(\eta)|\Psi(\eta)\rangle_I \label{eomip}\ee  where $H_I(\eta)$ is the interaction Hamiltonian in the interaction picture. In terms of the coefficients $C_n(\eta)$  eqn. (\ref{eomip}) becomes
   \be \frac{d\,C_n(\eta)}{d\eta}  = -i \sum_m C_m(\eta) \langle n|H_I(\eta)|m\rangle \,, \label{ecns}\ee it is convenient to separate the diagonal matrix elements, that represent \emph{local contributions}  from those that represent transitions and are associated with non-local self-energy corrections, writing
    \be \frac{d\,C_n(\eta)}{d\eta}  = -i C_n(\eta)\langle n|H_I(\eta)|n\rangle -i \sum_{m\neq n} C_m(\eta) \langle n|H_I(\eta)|m\rangle \,. \label{ecnsoff}\ee
    Although this equation is exact, it yields   an infinite hierarchy of simultaneous equations when the Hilbert space of states $|n\rangle$ is infinite dimensional. However, progress is made by considering the transition between states connected by the interaction Hamiltonian at a given order in $H_I$:
consider the case when one state, say $|A\rangle$ couples to a set of states $|\kappa\rangle$, which couple back to $|A\rangle$ via $H_I$, to lowest order in the interaction the system of equation closes in the form
 \bea \frac{d\,C_A(\eta)}{d\eta} & = & -i   \langle A|H_I(\eta)|A\rangle \, C_A(\eta)-i \sum_{\kappa \neq A} \langle A|H_I(\eta)|\kappa\rangle \,C_\kappa(\eta)\label{CA}\\
\frac{d\,C_\kappa(\eta)}{d\eta}& = & -i \, C_A(\eta) \langle \kappa|H_I(\eta) |A\rangle \label{Ckapas}\eea where the $\sum_{\kappa \neq A}$ is over all the intermediate states coupled to $|A\rangle$ via $H_I$ representing transitions.

Consider the initial value problem in which at time $\eta=\eta_0$ the state of the system is given by $|\Psi(\eta=\eta_0)\rangle = |A\rangle$ so that \be C_A(\eta_0)= 1 ~~;~~ C_{\kappa\neq A}(\eta=\eta_0) =0 \,,\label{initial}\ee  solving (\ref{Ckapas}) and introducing the solution into (\ref{CA}) we find \bea  C_{\kappa}(\eta) & = &  -i \,\int_{\eta_0}^{\eta} \langle \kappa |H_I(\eta')|A\rangle \,C_A(\eta')\,d\eta' \label{Ckapasol}\\ \frac{d\,C_A(\eta)}{d\eta} & = & -i   \langle A|H_I(\eta)|A\rangle \, C_A(\eta) - \int^{\eta}_{\eta_0} \Sigma_A(\eta,\eta') \, C_A(\eta')\,d\eta' \label{intdiff} \eea where\footnote{In ref.\cite{boyhol} it is proven that in Minkowski space-time the retarded self-energy in the single particle propagator is given by $i\Sigma$.} \be \Sigma_A(\eta,\eta') = \sum_{\kappa \neq A} \langle A|H_I(\eta)|\kappa\rangle \langle \kappa|H_I(\eta')|A\rangle \,. \label{sigma} \ee In eqn. (\ref{Ckapas}) we have not included the diagonal term as in (\ref{CA})\footnote{These diagonal terms represent local self-energy insertions in the propagators of the intermediate states, hence higher orders in the perturbative expansion.}, it is clear from (\ref{Ckapasol}) that with the initial condition (\ref{initial}) the amplitude of $C_{\kappa}$ is of $\mathcal{O}(H_I)$ therefore a diagonal term would effectively lead to higher order contributions to (\ref{intdiff}). The integro-differential equation  (\ref{intdiff}) with \emph{memory} yields a non-perturbative solution for the time evolution of the amplitudes and probabilities, which simplifies in the case of weak couplings. In perturbation theory   the time evolution of $C_A(\eta)$ determined by eqn. (\ref{intdiff}) is \emph{slow} in the sense that
the time scale is determined by a weak coupling kernel $\Sigma_A$, hence an approximation in terms of an expansion in derivatives of $C_A$ emerges as follows: introduce
\be W (\eta,\eta') = \int^{\eta'}_{\eta_0} \Sigma_A(\eta,\eta'')d\eta'' \label{Wo}\ee so that \be \Sigma_A(\eta,\eta') = \frac{d}{d\eta'}\,W (\eta,\eta'),\quad W (\eta,\eta_0)=0. \label{rela} \ee Integrating by parts in eq.(\ref{intdiff}) we obtain \be \int_{\eta_0}^{\eta} \Sigma_A(\eta,\eta')\,C_A(\eta')\, d\eta' = W (\eta,\eta)\,C_A(\eta) - \int_{\eta_0}^{\eta} W (\eta,\eta')\, \frac{d}{d\eta'}C_A(\eta') \,d\eta'. \label{marko1}\ee The second term on the right hand side is formally of \emph{higher order} in $H_I$, integrating by parts successively   yields  a systematic approximation scheme as discussed in ref.\cite{boyhol}.

 Therefore to leading order in the interaction we find
 \be C_A(\eta) = e^{-\int^{\eta}_{\eta_0}\widetilde{W} (\eta',\eta')\, d\eta'}  , \quad \widetilde{W} (\eta',\eta')= i   \langle A|H_I(\eta')|A\rangle + \int_{\eta_0}^{\eta'} \Sigma_A(\eta',\eta^{''}) d\eta^{''}\,. \label{dssolu} \ee

Following ref.\cite{boywwds}    we introduce the \emph{real quantities} $\mathcal{E}_A(\eta)\,;\,\Gamma_A(\eta)$ as
 \be  i   \langle A|H_I(\eta')|A\rangle +\int^{\eta'}_{\eta_0} \Sigma_A(\eta',\eta'')d\eta'' \equiv i\,\mathcal{E}_A(\eta')+ \frac{1}{2}~\Gamma_A(\eta') \label{realima}\ee in terms of which
 \be  C_A(\eta) = e^{-i\int^{\eta}_{\eta_0}\mathcal{E}_A(\eta') d\eta'}~ e^{-\frac{1}{2}\int^{\eta}_{\eta_0}\Gamma_A(\eta') d\eta'} \label{caofeta}\ee When the state $A$ is a single particle state, radiative corrections to the mass are extracted from $\mathcal{E}_A$ and
 \be \Gamma_A(\eta) = -  \frac{d}{d\eta}\ln\Big[|C_A(\eta)|^2\Big] \label{decarate} \ee is identified as a (conformal) time dependent decay rate.

 \vspace{2mm}

  \textbf{Extracting the mass:} In Minkowski space-time for $|A\rangle = |1_{\vk}\rangle$ a single particle state of momentum $\vk$, $\mathcal{E}_{1_{\vk}}$ includes the self-energy correction to the mass of the particle\cite{boyhol,boywwds,qoptics}. Consider adding a mass counterterm to the Hamiltonian density, in terms of the spatial Fourier transform of the fields it is given by
  \be H_{ct} = \frac{\delta M^2}{2} \sum_{\vec{k}} \pi_{\vk}\,\pi_{-\vk} \label{hctex}\ee the matrix element
  \be \langle 1^{\pi}_{\vk} | H_{ct} |1^{\pi}_{\vk} \rangle = \delta M^2 \,|g_\pi(\eta)|^2 \,, \label{mtxelex}\ee hence it is clear that only the imaginary part of $\widetilde{W}$ can be interpreted as a mass term, thus only the imaginary part of $\Sigma_{1_{\vk}}$ contributes to the mass. However, the non-local nature of $\Sigma_{1_{\vk}}$ also includes transient behavior from the initial state preparation thus a mass term must be isolated in the asymptotic long time limit when transient phenomena has relaxed. Last but not least momentum dependence can mask a constant mass term, which can only be identified in the long wavelength limit. In particular in refs.\cite{boyhol,boywwds} it is shown that  in Minkowski space time (see appendix)
  \be \mathrm{Im} \int^{t\rightarrow \infty}_0 \Sigma_{1_{\vk}}(t,t')dt' = \delta E_{1_{\vk}} \label{minki}\ee where $\delta E_{1_{\vk}}$ is the second order correction to the energy of a single particle state with momentum $\vk$ obtained in quantum mechanical perturbation theory (see also the appendix).

The program of renormalized perturbation theory begins by writing the free field part of the Lagrangian in terms of the renormalized mass and introducing a counterterm in the interaction Lagrangian so that it cancels the radiative corrections to the mass from the self-energy. Namely the counterterm in the interaction Lagrangian is fixed by requiring that   $\mathcal{E}_{1\vec{k}}(\eta') = 0 $, in the long time limit $\eta' \rightarrow 0^-$ and in the long-wavelength limit. Therefore as per the discussion above we   extract the mass term from the condition
 \be \mathcal{E}_{1\vec{k}}(\eta') =\langle 1_{\vec{k}}|H_I(\eta')|1_{\vec{k}}\rangle + \int_{\eta_0}^{\eta'} \mathrm{Im}\Big[\Sigma_1(k;\eta',\eta^{''})\Big] d\eta^{''} = 0\,  \label{renmass}\ee in the long wavelength limit.

 In Minkowski space time, the condition (\ref{renmass}) is tantamount to requiring that the (real part of the) pole in the propagator be at the physical mass\cite{boyhol} and is equivalent to the counterterm approach described in section (\ref{subsec:counter}).   In the appendix we carry out this program and show explicitly how the Wigner-Weisskopf approach reproduces the results in Minkowski space time obtained in section (\ref{sec:minkowski}) and how the mass is reliably extracted in the long time, long wavelength limit.

  We   implement the \emph{same strategy} to obtain the self-consistent radiatively generated mass in de Sitter space time where equation (\ref{renmass}) will determine the \emph{self-consistent condition} for the mass.

In the mass terms in the Lagrangian (\ref{lagrads}) we implement the counterterm method by introducing the renormalized masses $\mathcal{M}^2_{\chi,\pi}$ that include the radiative corrections, and writing
\be -\frac{M^2_\chi }{2\,H^2\,\eta^2}\,\chi^2- \frac{M^2_\pi }{2\,H^2\,\eta^2}\,\pi^2 \equiv -\frac{\mathcal{M}^2_\chi }{2\,H^2\,\eta^2}\,\chi^2- \frac{\mathcal{M}^2_\pi }{2\,H^2\,\eta^2}\,\pi^2 -\mathcal{L}_{ct} \label{ctdsdef}\ee leading to the counterterm \emph{Hamiltonian}
\be H_{ct} =  \frac{1}{2\,H^2\,\eta^2}~ \int d^3x   \Bigg[  \big(M^2_\chi -\mathcal{M}^2_\chi \big)   \,\chi^2 + \big(M^2_\pi - \mathcal{M}^2_\pi \big)  \,\pi^2 \Bigg] \label{Hctds}\ee included in the interaction Hamiltonian $H_I(\eta)$, and redefining
\be \Delta_\chi = \frac{\mathcal{M}^2_\chi}{3H^2}~~;~~\Delta_\pi = \frac{\mathcal{M}^2_\pi}{3H^2}\,.\label{newdeltas}\ee

In what follows we assume that $\Delta_{\chi,\pi} \ll 1$, therefore the leading order contributions arise from  poles in $\Delta_{\chi,\pi}$ as a result of the strong infrared divergences of minimally coupled light fields.

The contributions from diagrams like those of fig. (\ref{fig:selfenergy}, (a),(b),(c)) are cancelled by the tadpole condition (\ref{tadpoleds}). For the $\pi - \chi$-fields respectively we find
\be \langle 1^\pi_{\vk}|H_I(\eta)|1^\pi_{\vk} \rangle = \frac{|g_{\pi}(k,\eta)|^2 }{ H^2\,\eta^2}\, \, \Bigg[ \frac{\lambda}{2}~ \Big(\frac{J}{H^2}+3 \mathcal{I}_\pi + \mathcal{I}_\chi \Big)-\frac{\mathcal{M}^2_\pi}{H^2}\Bigg]   \,.\label{diagMEpi}\ee where $\mathcal{I}_{\chi,\pi}$ are given by
eqns.(\ref{Isdef}) with the redefined $\Delta_{\chi,\pi}$ given by (\ref{newdeltas}).

The non-local contribution is given by (see \cite{boywwds})
 \be \Sigma_{\pi}(k;\eta;\eta') = \frac{\lambda^2\,\sigma^2_0}{H^2\,\eta\,\eta'} \,g^*_{\pi}(k;\eta)g_{\pi}(k;\eta')\,\int \frac{d^3q}{(2\pi)^3}\,g_{\chi}(q;\eta)g^*_{\chi}(q;\eta')g_{\pi}(|\vec{q}-\vk|;\eta)
 g^*_{\pi}(|\vec{q}-\vk|;\eta')\,, \label{selfpids}\ee  For $\Delta_{\pi,\chi} \sim 0$ the integral features infrared divergences in the regions $q\sim 0;|\vec{q}-\vk|\sim 0$ which are manifest as poles in $\Delta_{\pi,\chi}$\cite{boywwds}. These regions are isolated following the procedure of ref.\cite{boywwds} and the poles in $\Delta_{\pi,\chi}$ can be extracted unambiguously. To leading order in these poles we find
 \be \Sigma_{\pi}(k;\eta;\eta') = \frac{\lambda^2\,\sigma^2_0}{8\pi^2\,H^2\,(\eta\,\eta')^2} \,g^*_{\pi}(k;\eta)g_{\pi}(k;\eta')\Bigg[ \frac{g_{\pi}(k;\eta)
 g^*_{\pi}(k;\eta')}{\Delta_\chi}+\frac{g_{\chi}(k;\eta)
 g^*_{\chi}(k;\eta')}{\Delta_\pi}
 \Bigg] \label{sigpipoles}
 \ee
 As discussed in detail in ref.\cite{boywwds} the poles originate in the emission and absorption of superhorizon quanta and arise from the integration of a band of superhorizon wavevectors $0\leq q \leq \mu_{ir} \rightarrow 0$ (see ref.\cite{boywwds} for details).

 As per the discussion in Minkowski space-time, a vanishing mass for a Goldstone boson after radiative correction requires that the tadpole terms in (\ref{diagMEpi}) be \emph{exactly} cancelled by the non-local self-energy contribution in the long-time, long wavelength limit. In particular the poles in $\Delta_{\chi,\pi}$ in (\ref{diagMEpi}) must be exactly cancelled by similar poles in $\Sigma_\pi$ (\ref{sigpipoles}). Therefore, to leading order in $\Delta_{\pi,\chi}$ we can set $\Delta_\pi = \Delta_\chi =0$, namely $\nu_{\pi,\chi}= 3/2$ in the mode functions $g_{\pi,\chi}$ given by (\ref{gnu}), whence it follows that to leading order in $\Delta_{\pi,\chi}$
  \be \Sigma_{\pi}(k;\eta;\eta') = \frac{\lambda^2\,\sigma^2_0}{8\pi^2\,H^2} \,\frac{|g(k;\eta)|^2|g (k;\eta')|^2}{(\eta\,\eta')^2}\Big[\frac{1}{\Delta_\pi}+ \frac{1}{\Delta_\chi}\Big]\Big[1+\mathcal{O}(\Delta_\pi,\Delta_\chi)+\cdots\Big]\,, \label{selfleadds}\ee where
  \be g(k;\eta) =  -\frac{1}{2}
\sqrt{-\pi \eta}\,H^{(1)}_{\frac{3}{2}}(-k\eta)\,. \label{g32}\ee

  Therefore, to leading order in poles in $\Delta_{\chi,\pi}$,  $\Sigma_\pi(k;\eta;\eta')$ is \emph{real} and \emph{does not contribute to the radiatively generated $\pi$ mass }.

  Therefore, to leading order in the poles in $\Delta_{\pi,\chi}$ the self-consistent condition that determines the mass, eqn. (\ref{renmass}) becomes
  \be\langle 1^\pi_{\vk}|H_I(\eta)|1^\pi_{\vk} \rangle =0 \,.\label{finicondi}\ee

  This observation is important: unlike Minkowski space time where the diagram (\ref{fig:selfenergy}-(f)) \emph{cancels} the local tadpole contributions, in de Sitter space time the similar diagram \emph{cannot} cancel the local contributions because the leading infrared divergences yield a real contribution whereas the tadpoles yield a purely imaginary contribution as befits a mass insertion. Therefore, the self-consistent mass is obtained solely from the local tadpole terms which determine the mean-field contribution. This validates     the results of \cite{serreau,prokossb} which rely solely on the mean field approximation (which is exact only in the strict $N\rightarrow \infty$ limit).

   \emph{Assuming} spontaneous symmetry breaking so that eqn. (\ref{tadpoleds}) is fulfilled with $\sigma_0 \neq 0$, namely
  \be \frac{J}{H^2} = -3\mathcal{I}_\chi-\mathcal{I}_\pi \, , \label{ssb22}\ee it follows that
  \be \frac{\mathcal{M}^2_\pi}{H^2} = \frac{\lambda }{8\pi^2}\Big[\frac{1}{\Delta_\pi}-\frac{1}{\Delta_\chi} \Big] \,.\label{masscondids}\ee

For the $\chi$ field we find the following contributions,

 \be \langle 1^\chi_{\vk}|H_I(\eta)|1^\chi_{\vk} \rangle = \frac{|g_{\chi}(k,\eta)|^2 }{ H^2\,\eta^2}\, \, \Bigg[ \frac{\lambda}{2}~ \Big(\frac{J}{H^2}+2\,\frac{\sigma^2_0}{H^2}+3 \mathcal{I}_\chi + \mathcal{I}_\pi \Big)-\frac{\mathcal{M}^2_\chi}{H^2}\Bigg]   \,.\label{diagMEchi}\ee

 where $\mathcal{I}_{\chi,\pi}$ are given by eqn. (\ref{Isdef}) and for $\Sigma_\chi(k;\eta,\eta')$ we find
 \bea \Sigma_{\chi}(k;\eta;\eta') & = &  \frac{\lambda^2\,\sigma^2_0}{2\,H^2\,\eta\,\eta'}   g^*_{\chi}(k;\eta)g_{\chi}(k;\eta') \int  \frac{d^3q}{(2\pi)^3}   \Bigg[9\, g_{\chi}(q;\eta)g^*_{\chi}(q;\eta')g_{\chi}(|\vec{q}-\vk|;\eta)
 g^*_{\chi}(|\vec{q}-\vk|;\eta')   \nonumber \\ & + &  g_{\pi}(q;\eta)g^*_{\pi}(q;\eta')g_{\pi}(|\vec{q}-\vk|;\eta)
 g^*_{\pi}(|\vec{q}-\vk|;\eta') \Bigg]\,. \label{selfchids}\eea Extracting the poles in $\Delta_{\pi,\chi}$ the leading order result is given by
 \be \Sigma_{\chi}(k;\eta;\eta') = \frac{\lambda^2\,\sigma^2_0}{8\pi^2\,H^2\,(\eta\,\eta')^2} \,g^*_{\chi}(k;\eta)g_{\chi}(k;\eta')\Bigg[ \frac{g_{\pi}(k;\eta)
 g^*_{\pi}(k;\eta')}{\Delta_\pi}+9\,\frac{g_{\chi}(k;\eta)
 g^*_{\chi}(k;\eta')}{\Delta_\chi}
 \Bigg] \label{sigchipoles}
  \ee
  Again, just as for the $\pi$ field above, to leading order in the poles in $\Delta_{\pi,\chi}$ we can set
  $\Delta_\pi = \Delta_\chi =0$, namely $\nu_{\pi,\chi}=3/2$ in the mode functions $g_{\pi,\chi}$, leading to
  \be \Sigma_{\chi}(k;\eta;\eta') = \frac{\lambda^2\,\sigma^2_0}{8\pi^2\,H^2\,} \,\frac{|g(k;\eta)|^2|g(k;\eta')|^2}{(\eta\,\eta')^2}\Bigg[ \frac{1}{\Delta_\pi}+ \frac{9}{\Delta_\chi}
 \Bigg] \label{sigchilead}
  \ee  where $g(k;\eta)$ is given by eqn. (\ref{g32}).

  The result is that to leading order in the poles, \emph{both} $\Sigma_{\pi,\chi}$ are \emph{real} and \emph{do not contribute to the radiatively generated masses} but will contribute to the \emph{decay} of the single particle excitations discussed below (see section \ref{subsec:decay}).

  Therefore, \emph{assuming} spontaneous symmetry breaking so that the condition (\ref{ssb22}) holds we find that
  \be \frac{\mathcal{M}^2_\chi}{H^2} =   \frac{\lambda \, \sigma^2_0}{H^2}\,. \label{masachids}\ee

  Now identifying self-consistently the masses in the definition (\ref{newdeltas}) with $\mathcal{M}_{\pi,\chi}$,   and defining
  \be \varepsilon = \sqrt{\frac{\lambda}{24\pi^2}}~~;~~ \Delta_\pi = \varepsilon \delta_\pi~~;~~\Delta_\chi = \frac{\lambda}{3}\,\frac{\sigma^2_0}{H^2} \equiv \varepsilon \delta_\chi \label{defsds}\ee   equation (\ref{masscondids}) becomes
  \be \delta_\pi = \frac{1}{\delta_\pi}- \frac{1}{\delta_\chi} \label{delpieq}\ee with the (positive) solution \be \delta_\pi = \frac{1}{2\delta_\chi}\Big[\sqrt{1+4\delta^2_\chi}-1 \Big]\label{delpisol}\ee the negative root would lead to an instability and an uncontrollable infrared divergence in the loop integrals which would not yield a self-consistent solution.

\begin{figure}[ht!]
\begin{center}
\includegraphics[height=4in,width=4in,keepaspectratio=true]{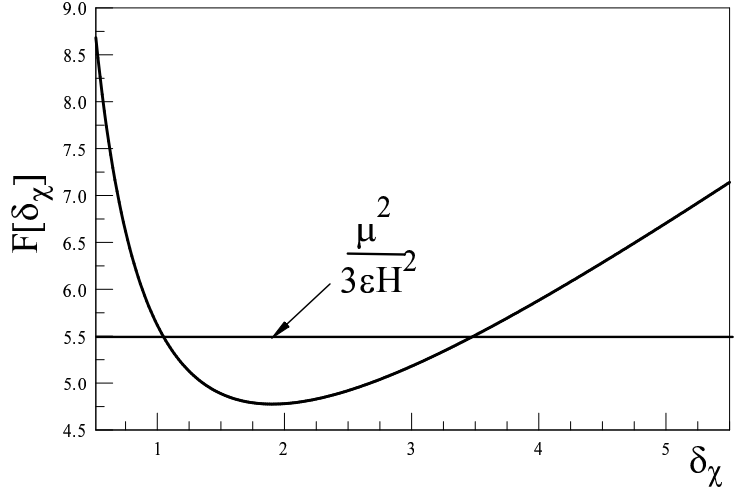}
\caption{$F[\delta_\chi]$ vs. $\delta_\chi$ and its intersection with $\mu^2/3\varepsilon H^2$. The function features a minimum at $\delta_{\chi,min} = 1.906\cdots$ with $F[\delta_{\chi,min}]= 4.77614\cdots$. The value of $\delta_\pi(\delta_{\chi,min})= 0.772\cdots$. }
\label{fig:ssbsol}
\end{center}
\end{figure}

  Now we are in position to understand whether spontaneous symmetry breaking does occur. The condition (\ref{ssb22}) is
  \be \frac{\sigma^2_0}{H^2} = \frac{\mu^2}{\lambda H^2} - \frac{3}{8\pi^2\,\Delta_\chi}-\frac{1}{8\pi^2\,\Delta_\pi} \neq 0\label{condi2}\ee which when written in terms of the definitions (\ref{defsds}) and using (\ref{delpisol})  becomes
  \be F[\delta_\chi]\equiv \delta_\chi +\frac{1}{2\delta_\chi}\Big[7+ \sqrt{1+4\delta^2_\chi} ~ \Big]= \frac{\mu^2}{3\varepsilon H^2} \label{funcondi} \ee The function $F[\delta_\chi]$ and its intersection with $\mu^2/3\varepsilon H^2$ is displayed in fig. (\ref{fig:ssbsol}).

As shown in fig. (\ref{fig:ssbsol}), $F[\delta_\chi]$ features a minimum at $\delta_{\chi,min}= 1.906\cdots$ at which $F[\delta_{\chi,min}]= 4.77614\cdots$, therefore there are symmetry breaking solutions for
\be \frac{\mu^2}{3\varepsilon H^2} > 4.77614\cdots \label{ssbsols}\ee this condition can be written in a more illuminating manner as
\be T_H < T_c ~~;~~T_H = \frac{H}{2\pi} ~~;~~ T_c = \frac{\mu}{2.419\cdots\,\lambda^{1/4}} = \frac{\lambda^{1/4}\,v }{2.419\cdots}  \label{crittemp}\ee where $T_H$ is the Hawking temperature of de Sitter space time\footnote{In comoving time $t$, the mode functions $g_\pi,g_\chi$ are functions of $\eta = -e^{-Ht}/H$ therefore   \emph{periodic} in imaginary time $\tau=it$ with period $\beta=2\pi/H=1/T_H$. See \cite{boyprem}.} and $v=\mu/\sqrt{\lambda}$ is the tree level vacuum expectation value (minimum of the tree level potential). From eqn. (\ref{delpieq}) it follows that
 \be \frac{\delta_\chi}{\delta_\pi} = \frac{1+\sqrt{1+4\delta^2_\chi}}{2} \label{ratio}\ee and $\delta_\chi > 1.906\cdots$, therefore in the broken symmetry phase we find that
 \be \frac{\delta_\chi}{\delta_\pi} \simeq \delta_\chi +\frac{1}{2} ~~\mathrm{for}~~T_H < T_c \,, \label{aprossb}\ee   in the spontaneously broken phase. At weak coupling, for $\mu^2 \gg 3 \varepsilon H^2$ (but $\mu^2 \ll H^2$ for consistency ) we find that
\be \mathcal{M}_\chi \simeq |\mu| + a \,\lambda^{1/4}\,H ~~;~~ \mathcal{M}_\pi = b   \,\lambda^{1/4}\,H \, \label{wcmasses}\ee where $a,b$ are positive constants.

For $T_H > T_c$   the unbroken symmetry solution $\sigma_0=0$ is the only solution of the tadpole condition (\ref{tadpoleds}). In this case we find
\be \frac{\mathcal{M}^2_\pi}{H^2}= \frac{\lambda}{2}~ \Big(\frac{J}{H^2}+3 \mathcal{I}_\pi + \mathcal{I}_\chi \Big) \label{thgtcpi}\ee
\be  \frac{\mathcal{M}^2_\chi}{H^2} = \frac{\lambda}{2}~ \Big(\frac{J}{H^2}+3 \mathcal{I}_\chi + \mathcal{I}_\pi \Big)  \label{thgtcchi}\ee subtracting (\ref{thgtcchi}) from (\ref{thgtcpi}) we find
\be \delta_\pi -  \delta_\chi = \frac{1}{\delta_\pi} - \frac{1}{\delta_\chi} \,, \label{diffa}\ee if $\delta_\pi > (<) \, \delta_\chi$ the left hand side is positive (negative) but the right hand side is negative (positive), therefore the only solution is
\be \delta_\pi = \delta_\chi  =  \frac{\mu^2}{12\varepsilon H^2}\Bigg[\sqrt{1+\Big(\frac{12\varepsilon H^2}{\mu^2} \Big)^2}-1 \Bigg] \,.\label{thgtcdels}\ee Inserting this result in (\ref{thgtcchi}) we find for $T_H > T_c$
\be \mathcal{M}_\pi = \mathcal{M}_\chi = \frac{\mu^2}{4}\Bigg[\sqrt{1+0.701\Big(\frac{  T^2_H}{T^2_c} \Big)^2}-1 \Bigg] \,,\label{thgtcmass}\ee as expected $ \mathcal{M}_{\chi}=\mathcal{M}_{\pi}$ if the symmetry is unbroken.

\subsection{A first order phase transition:}

Fig. (\ref{fig:ssbsol}) shows that for $T_H < T_c$ there are \emph{two solutions} of the equation that determines symmetry breaking and the question arises: which of the two solutions describes the broken symmetry phase?. The answer is gleaned by analyzing the weak coupling limit $\varepsilon \rightarrow 0$ ($\lambda \rightarrow 0)$. In this limit the left most intersection in fig.(\ref{fig:ssbsol}) corresponds to the solution
\be \delta^{(-)}_\chi \simeq 12\,\varepsilon \,\frac{H^2}{\mu^2} \Rightarrow \mathcal{M}^2_\chi \simeq \frac{\lambda}{2\pi^2} \frac{H^4}{\mu^2} \stackrel{\lambda \rightarrow 0}{\rightarrow 0}  \label{solumin}\ee whereas the right-most intersection corresponds to the solution
\be  \delta^{(+)}_\chi \simeq \frac{\mu^2}{3\varepsilon H^2} \Rightarrow \mathcal{M}^2_\chi \simeq \mu^2 ~~;~~ \mathcal{M}^2_\pi \simeq \varepsilon\,H^2 \rightarrow 0 \label{soluplus}\ee Obviously the solution $\delta^{(+)}_\chi$ is the correct one since for $\lambda \rightarrow 0$  the expectation value $\lambda \sigma_0 = \mu^2 $ the     loop corrections vanish and the mass of the $\chi,\pi$ fields should be the tree level ones namely $\mathcal{M}^2_\chi = \mu^2,\mathcal{M}^2_\pi = 0$ respectively. However, as $\varepsilon H^2$ increases beyond the critical value at which $\mu^2/3\varepsilon H^2 = F[\delta_{\chi,min}]$   there is no available symmetry breaking solution and this occurs for a non-vanishing value of $\sigma_0$ signaling a \emph{first order phase transition} at   $T_H =   T_c$ given by (\ref{crittemp}). The value of the order parameter at $T_H=T_c$ is given by
\be \sigma_{0c} \simeq 0.61\, \frac{H}{\lambda^{1/4}} \,.\label{critop}\ee

These results are in general agreement with those of ref.\cite{prokossb}. The first order nature of the phase transition can also be understood within the context of the infrared divergences: if the transition (as a function of coupling or $T_H$) were of second order, then at the critical point the masses of both $\chi,\pi$ fields must necessarily vanish, but the vanishing of the masses would lead to strong infrared divergences. Therefore a first order transition  with a finite mass (correlation length) and a jump in the order parameter is a natural consequence of the strong infrared behavior of minimally coupled nearly massless fields in de Sitter space-time. The infrared singularities are self-consistently relieved by the radiative generation of a mass at the expense of turning the phase transition into first order.

\subsection{Large N limit}
The above results can be simply generalized to the $O(N)$ case where the $\pi$-fields form an $O(N-1)$ multiplet. Now the tadpole condition   becomes
  \be \langle \chi \rangle = 0 \Rightarrow \frac{\lambda\,a\,\sigma_0}{2\,\eta^2}\Big[\frac{J}{H^2}+3\mathcal{I}_{\chi}+(N-1)\mathcal{I}_{\pi}\Big] =0 \label{tadpoledsN}\ee and the $\chi,\pi$ masses become
 \be  \frac{\mathcal{M}^2_\pi}{H^2}  =   \frac{\lambda}{2}~ \Big[\frac{J}{H^2}+ (N+1) \mathcal{I}_\pi + \mathcal{I}_\chi \Big]  \label{masspiN}  \ee
 \be \frac{\mathcal{M}^2_\chi}{H^2} =   \frac{\lambda}{2}~ \Big[2\frac{\sigma^2_0}{H^2}+\frac{J}{H^2}+ (N-1) \mathcal{I}_\pi + 3\mathcal{I}_\chi \Big]\,.  \label{masschiN}\ee In the strict $N\rightarrow \infty$ limit these equations simplify to
 \bea && \sigma_0 \Big[\frac{J}{H^2} +N\mathcal{I}_{\pi}\Big] =0  \label{largeNtad} \\
 &&  \frac{\mathcal{M}^2_\pi}{H^2}  =   \frac{\lambda}{2}~ \Big[\frac{J}{H^2}+ N \mathcal{I}_\pi  \Big] \label{largeNmasspi}\\
&& \frac{\mathcal{M}^2_\chi}{H^2} =   \frac{\lambda}{2}~ \Big[2\frac{\sigma^2_0}{H^2}+\frac{J}{H^2}+ N \mathcal{I}_\pi   \Big]\,, \label{largeNmasschi}\eea with $\mathcal{I}_{\pi,\chi}$ given by eqn.(\ref{deltas})
and self-consistently $\Delta_{\pi,\chi} = \mathcal{M}^2_{\pi,\chi}/3H^2$. Clearly, eqns. (\ref{largeNtad},\ref{largeNmasspi}) lead  to conclude that the only symmetry breaking solution corresponds to $\mathcal{M}^2_\pi =0$ but this is obviously in contradiction with the self-consistent solution because of the infrared singularity in $\mathcal{I}_\pi \propto 1/\mathcal{M}^2_\pi$. Therefore, the only available solution of (\ref{largeNtad})  that is also self-consistent and infrared finite must be the unbroken symmetry solution $\sigma_0=0$ which results in equal masses for $\chi,\pi$ fields. Thus in the strict $N\rightarrow \infty$, neglecting the $1/N$ corrections the $O(N)$ symmetry \emph{cannot} be spontaneously broken because of the strong infrared effects. This is the conclusion of ref.\cite{serreau}.
 However the analysis presented above for finite $N$, and in particular for $N=2$ suggests that this conclusion  holds \emph{only} in the strict $N\rightarrow \infty$ limit but for any finite $N$ \emph{there is spontaneous symmetry breaking}, along with infrared radiatively induced masses for the Goldstone fields without contradicting Goldstone's theorem, but the transition is first order as a consequence of infrared divergences.

 \vspace{2mm}

 \subsection{Decay of $\pi,\chi$ particles:} \label{subsec:decay}

 As discussed above the non-local self-energies $\Sigma_{\pi,\chi}(k;\eta,\eta')$ are real and do not contribute to the mass to leading order in $\Delta_{\pi,\chi}$, however they determine the \emph{decay} of single particle states as described in ref.\cite{boywwds}. We now focus in obtaining the decay amplitudes arising from these contributions.  Using the relations given by eqns. (\ref{masachids}-\ref{defsds}) to leading order in poles in $\Delta_{\pi,\chi}$ the one loop results  (\ref{selfleadds},\ref{sigchilead}) can be written as
 \bea \Sigma^{(1)}_{\pi}(k;\eta;\eta') &  = &  \frac{3\,\lambda }{8\pi^2} \,\frac{|g(k;\eta)|^2|g (k;\eta')|^2}{(\eta\,\eta')^2}\Big[1+\frac{\delta_\chi}{\delta_\pi}\Big] \label{sigpifi}\\
 \Sigma^{(1)}_{\chi}(k;\eta;\eta') &  = &  \frac{27\,\lambda }{8\pi^2} \,\frac{|g(k;\eta)|^2|g (k;\eta')|^2}{(\eta\,\eta')^2}\Big[1+\frac{\delta_\chi}{9\delta_\pi} \Big] \label{sigchifi}\,.
 \eea

 Thus formally the real part of the single particles self-energy are of $\mathcal{O}(\lambda)$.

\begin{figure}[ht!]
\begin{center}
\includegraphics[height=4in,width=4in,keepaspectratio=true]{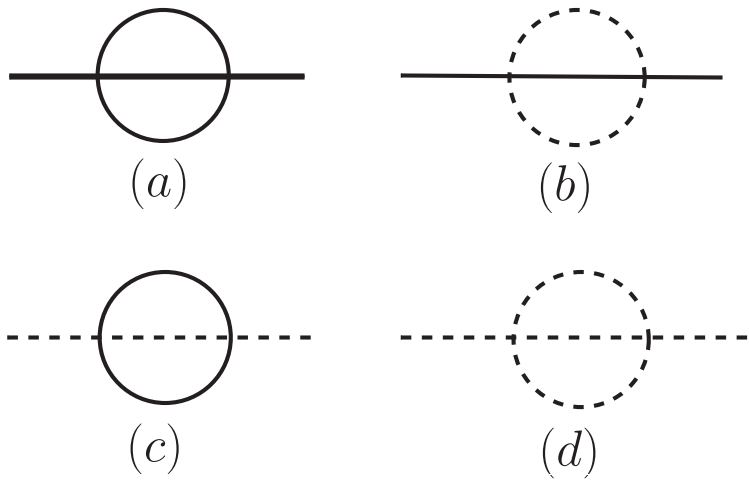}
\caption{Two loops contributions to $\Sigma_\pi$ (a,b) and $\Sigma_\chi$ (c,d). Solid lines $=\pi$, dashed lines $=\chi$  . }
\label{fig:twoloops}
\end{center}
\end{figure}

  In ref.(\cite{boywwds}) it was found that quartic self-interactions with strength $\lambda$ yield \emph{two loops} self-energies that are \emph{also} of $\mathcal{O}(\lambda)$ as a consequence of infrared divergences that are manifest as \emph{second order} poles in $\Delta$. Implementing the ``infrared rules'' obtained in ref.\cite{boywwds} in the two loop diagrams for $\Sigma_{\pi,\chi}$ figs. (\ref{fig:twoloops}) (a,b) and (c,d) respectively we find the leading order two-loops contributions
  \bea \Sigma^{(2)}_{\pi}(k;\eta;\eta') &  =  & \frac{3\lambda}{16\pi^2}\,\frac{|g(k;\eta)|^2 |g(k;\eta')|^2}{(\eta\,\eta')^2}\,\Bigg[\frac{9}{\delta^2_\pi}+\frac{1}{\delta^2_\chi}+\frac{2}{\delta_\pi\delta_\chi} \Bigg] \label{sigpi2} \\\Sigma^{(2)}_{\chi} (k;\eta;\eta') &  =  & \frac{3\lambda}{16\pi^2}\,\frac{|g(k;\eta)|^2 |g(k;\eta')|^2}{(\eta\,\eta')^2}\,\Bigg[\frac{9}{\delta^2_\chi}+\frac{1}{\delta^2_\pi}+\frac{2}{\delta_\pi\delta_\chi} \Bigg]\label{sigchi2}
  \eea

From (\ref{realima}) we obtain the conformal time dependent single particle decay rates (\ref{decarate})
\be \frac{1}{2}\Gamma_{\pi,\chi}(k;\eta) =  \int^{\eta}_{\eta_0} \Sigma_\pi(k;\eta;\eta') d\eta' = \lambda ~ \mathcal{C}_{\pi,\chi}\,
 \,k\,\frac{\big|H^{(1)}_{ {3}/{2}}(z)\big|^2}{z}\int^{z_0}_z \frac{dz'}{z'}\,
\big|H^{(1)}_{ {3}/{2}}(z')\big|^2  ~~;~~z=-k\eta \,, \label{gammas}\ee with
\bea \mathcal{C}_\pi & = & \frac{3}{256}\Bigg[2\Big( 1+\frac{\delta_\chi}{\delta_\pi}\Big) + \Big( \frac{9}{\delta^2_\pi}+\frac{1}{\delta^2_\chi}+\frac{2}{\delta_\pi\delta_\chi} \Big) \Bigg]\label{Cpi}\\
\mathcal{C}_\chi & = & \frac{3 }{256}\Bigg[18\Big( 1+\frac{\delta_\chi}{9\delta_\pi}\Big) + \Big( \frac{9}{\delta^2_\pi}+\frac{1}{\delta^2_\chi}+\frac{2}{\delta_\pi\delta_\chi} \Big) \Bigg]\label{Cchi}
\eea

As discussed in ref.\cite{boyhol,boywwds} the decay $\pi \rightarrow \pi + \chi$ is a consequence of emission and absorption of superhorizon quanta, in the superhorizon limit $z \ll 1~;~z_0 \sim 1$ the integrals can be done simply\cite{boywwds}, leading to the results for the single particle amplitudes in the superhorizon limit
\be |C^{\pi,\chi}_{1_{\vk}}| \simeq e^{-\gamma_{\chi,\pi}(-k\eta)}~~;~~ \gamma_{\chi,\pi}(-k\eta) = \frac{2\lambda}{9\pi^2}\,\mathcal{C}_{\chi,\pi}\,\Bigg[ \frac{H}{k_{phys}(\eta)}\Bigg]^6 \,. \label{decaypichi}\ee

\begin{figure}[ht!]
\begin{center}
\includegraphics[height=4in,width=4in,keepaspectratio=true]{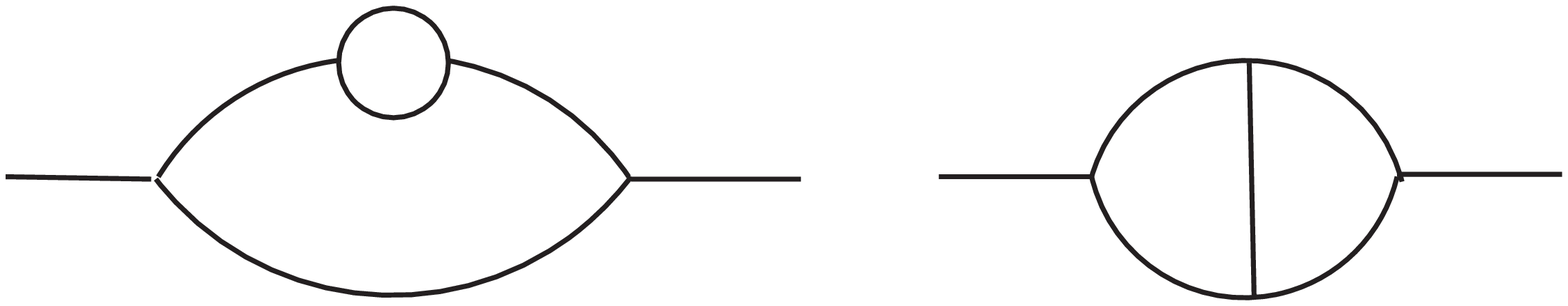}
\caption{Other two loops contributions to $\Sigma_{\pi,\chi}$  . }
\label{fig:othertwoloops}
\end{center}
\end{figure}

\textbf{Possible caveats:}
There are other two loops diagrams that have not been accounted for above.
The generic form of these diagrams are displayed in fig.(\ref{fig:othertwoloops}) (we have not displayed specific $\pi,\chi$ lines but just showed the generic form of the diagrams) and can be interpreted as a renormalization of the internal propagator and the vertex. Both of these diagrams are $\propto (\lambda \sigma_0/H)^4  \simeq \lambda^2 \Delta^2_\chi$, therefore \emph{if} the ``infrared rules'' of ref.\cite{boywwds} apply to these diagrams the two loops imply an infrared factor $\propto 1/\Delta^2_\chi; 1/\Delta^2_\pi ; 1/\Delta_\chi \Delta_\pi$, in which case the overall coupling dependence of these diagrams is $\propto \lambda^2$ and would be subdominant as compared to the two loop diagrams of fig. (\ref{fig:twoloops}). The possible caveat in this argument is that the rules to obtain the leading contributions in poles in $\Delta$ given in ref.\cite{boywwds} do not directly apply to the diagrams above because if the bubble that renormalizes the propagator in the first diagram dresses a line in which the wavevector  is within  an infrared band $0 < q < \mu_{ir}\rightarrow 0$, then both lines in this bubble are
within this band. This situation is not contemplated in the rules provided in ref.\cite{boywwds} which apply to the case when in a loop integral only one of the lines carries momenta within an infrared band whereas the other line carries a finite value of the momentum (even if superhorizon) (see the arguments in ref. \cite{boywwds}). Thus in absence of a sound proof that the diagrams in fig. (\ref{fig:othertwoloops}) are subleading, the result for the damping rate $\Gamma(k;\eta)$ given by eqn. (\ref{gammas}) should be taken as indicative. Nevertheless the analysis of symmetry breaking and the emerging conclusions on the mass generation of Goldstone bosons and the order of the transition are not affected by this possible caveat on the damping rate. Further study on the infrared aspects of diagrams in fig. (\ref{fig:othertwoloops}) is certainly worthy but beyond the scope of this article.

\section{Conclusions:}
Spontaneous symmetry breaking is an important ingredient in the inflationary paradigm. In this article we have studied (SSB) of continuous symmetry in an $O(2)$ model of scalar fields minimally coupled to gravity in de Sitter space time, focusing in particular on understanding whether Goldstone's theorem implies massless Goldstone bosons and trying to shed light on conflicting previous results\cite{serreau,prokossb} which implemented a local mean field approximation. We first revisited the general results of Goldstone's theorem in Minkowski space time highlighting the fact that it is through \emph{time translational invariance that the conservation of the Noether theorem guarantees massless Goldstone bosons}. We emphasized that in absence of time translational invariance Goldstone's theorem is much less stringent and does not rule out radiatively generated masses for Goldstone modes. We followed with an   analysis of the implementation of Goldstone's theorem at one loop level in Minkowski space-time by studying the self energies of Goldstone and Higgs-like modes, we showed that at one loop level the masslessness of the Goldstone boson is a consequence of a precise cancellation between local tadpole and non-local (in space-time) contributions, and analyzed in detail the implementation of Goldstone's theorem in the large N limit of an $O(N)$ scalar theory. These results paved the way towards a deeper understanding of Goldstone's theorem and its consequences in de Sitter cosmology.

Our conclusions are summarized as follows:

\begin{itemize}
\item In absence of a global time-like Killing vector Goldstone's theorem \emph{does not} imply massless Goldstone bosons when a continuous symmetry is spontaneously broken.

    \item We implemented a non-perturbative Wigner-Weisskopf method that allows to obtain the masses and decay widths of single particle states in a cosmological setting. Strong infrared behavior associated with light particles minimally coupled to gravity are treated in a self-consistent manner.

\item  Whereas in Minkowski space time at one loop level the masslessness of Goldstone modes in the broken symmetry phase is a consequence of a precise cancellation between tadpole and non-local (absorptive) contributions to the self energy, we find that in de Sitter space time   no such cancellation is possible.   Goldstone modes acquire a self-consistent radiatively generated mass resulting from the build-up of infrared singularities in self-energies. We find that  in a weak coupling the mass of the Goldstone modes is     $\mathcal{M}_\pi \propto \lambda^{1/4} H$, where $\lambda$ is the quartic coupling of the $O(2)$ theory.

\item We find a \emph{first order phase transition} between the broken and unbroken symmetry phase as a function of $T_H=H/2\pi$ the Hawking temperature of de Sitter space-time. For the $O(2)$ model we find (SSB) for $T_H < T_c = {\lambda^{1/4}\,v }/{2.419\cdots}  $   where $v$ is the tree level vacuum expectation value. For $T_H > T_c$ the symmetry is restored. The value of the order parameter at $T_H=T_c$ is  $\sigma_{0c} \simeq 0.61\, {H}/{\lambda^{1/4}}$. The first order nature of the transition and concomitant jump in the order parameter is a consequence of the strong infrared behavior of correlation functions: if the transition were second order both fields would be massless at $T_c$ leading to strong infrared singularities. Thus radiatively induced masses relieve the infrared singularities at the expense of a first order transition and a jump in the order parameter. These results are in qualitative agreement with those of ref.\cite{prokossb} and also confirm the validity of the local mean field approximation since the non-local radiative corrections do not contribute to the masses of either Goldstone or Higgs-like modes but only to their decay widths.

    \item In the strict $N \rightarrow \infty$ limit of an $O(N)$ scalar theory there is no possibility of (SSB) in agreement with the result of ref.\cite{serreau}, but (SSB) is available for any finite $N$. This result reconciles the conflicting conclusions of refs.\cite{serreau,prokossb}.

        \item The lack of a global time-like Killing vector prevents the existence of kinematic thresholds, as a result we find that Goldstone modes \emph{decay} into Goldstone and Higgs modes via the emission and absorption of superhorizon quanta. We have obtained the decay width of Goldstone modes in the superhorizon limit, the amplitude of single particle Goldstone modes $|C^{\pi}_{1_{\vk}}| \simeq e^{-\gamma_{\pi}(-k\eta)}$ where $\gamma_{ \pi}(-k\eta) \propto \lambda \,  \big(H/k_{phys}(\eta)\big)^6  $.

\end{itemize}

\vspace{2mm}

\textbf{Further Questions:} The discussion in section (\ref{sec:goldcosmo}) on the applicability and corollary  of Goldstone's Theorem in an expanding cosmology highlights the consequences of a covariant conservation law in a time dependent background geometry as contrasted with the strict conservation law in Minkowski space time and is  general for any cosmological background. Our study focused on de Sitter space time wherein infrared divergences associated with minimally coupled massless particles lead to the self-consistent generation of masses for Goldstone bosons as described above. There remains the very important question of whether Goldstone bosons acquire a mass in other cosmologies, for example during the radiation dominated stage, where the arguments on the time dependence of the background are valid but there may not be infrared divergences that lead to a self-consistent generation of mass as in de Sitter space time. A deeper understanding of this case certainly merits further study as it may yield to novel and unexpected phenomena in cosmology and is relegated to future work.

\acknowledgments The author acknowledges  support  by the NSF through   award   PHY-0852497.

\appendix
\section{Wigner-Weisskopf approach to Goldstone's theorem in Minkowski space time:}\label{appx:wwmink}

In Minkowski space time and for a single particle $\pi$ state of momentum $\vec{k}$ we need (see eqn. (\ref{dssolu}))
\be \widetilde{W}(t) = i\langle 1^{\pi}_{\vk}|H_I(t)|1^{\pi}_{\vk}\rangle + \int^t_0 \Sigma_\pi(\vk;t,t')\,dt' \label{mtxmink}\ee from which the total correction to the energy of a single particle state is obtained from the long time limit
 \be \mathcal{E}^{\pi}_{1\vec{k}} =\langle 1_{\vec{k}}|H_I(0)|1_{\vec{k}}\rangle + \int_0^{t\rightarrow\infty} \mathrm{Im}\Big[\Sigma_\pi(k;t,t') \Big]\,dt'  \,, \label{enemink}\ee Including the counterterm Hamiltonian in the interaction as described in section (\ref{subsec:counter}) leads to  the  requirement that in the long-wavelength limit
 \be \mathcal{E}^{\pi}_{1\vec{k}\rightarrow 0} = 0 \,.\label{countminki}\ee

The interaction Hamiltonian is read-off from the vertices in eqn. (\ref{potafter}) including the mass counterterm
\be H_{ct}= \frac{1}{2}\Big(M^2_\pi-\mathcal{M}^2_\pi\Big)~~;~~ M^2_\pi = \frac{\lambda}{2}\,J \,. \label{Hct}\ee The contribution $\langle 1^{\pi}_{\vk}|H_I(t)|1^{\pi}_{\vk}\rangle $ is recognized as the first order shift in the energy.

The tadpole condition eliminate the contributions from the tadpoles in figs.(\ref{fig:selfenergy} (a,b,c)) because the matrix element of the Hamiltonian between the vacuum and a single particle state vanish by dint of the tadpole condition. We find
\be \langle 1^{\pi}_{\vk}|H_I(0)|1^{\pi}_{\vk}\rangle = \frac{1}{2 \omega^{\pi}(k)}\Bigg[\Big(M^2_\pi-\mathcal{M}^2_\pi\Big)+  \frac{\lambda}{2}  \Big(I_\chi + 3 I_\pi \Big) \Bigg]\label{diag}\ee where $I_{\chi,\pi}$ are given by eqn. (\ref{tadints}). Upon using the tadpole condition assuming spontaneous symmetry breaking it follows that $J = -3I_\chi -I_\pi$ and (\ref{diag}) becomes
\be \langle 1^{\pi}_{\vk}|H_I(0)|1^{\pi}_{\vk}\rangle = \frac{1}{2 \omega^{\pi}(k)}\Bigg[ -\mathcal{M}^2_\pi + {\lambda}  \int \frac{d^3q}{(2\pi^3)}\Bigg(\frac{1}{2\omega^{\pi}(q)} - \frac{1}{2\omega^{\chi}(q)} \Bigg) \Bigg] \label{11HI}\ee
The   self-energy
\be \Sigma_\pi(k;t,t') = \sum_{\kappa\neq 1^{\pi}_{\vk}} \langle 1^{\pi}_{\vk}|H_I(t)|\kappa\rangle \langle\kappa|H_I(t')|1^{\pi}_{\vk}\rangle  = \sum_{\kappa\neq 1^{\pi}_{\vk}} |\langle 1^{\pi}_{\vk}|H_I(0)|\kappa\rangle|^2\,e^{i(\omega^{\pi}(k)-E_\kappa )(t-t')}\,, \label{sigpimin}\ee where the intermediate states $|\kappa \rangle = | 1^{\chi}_{\vq};1^{\pi}_{\vq+\vk}\rangle$  (see fig. (\ref{fig:selfenergy})-(f)).

Carrying out the time integral in (\ref{mtxmink}) in the long time limit we find
\be \int^{t\rightarrow \infty}_0\Sigma_\pi(k;t,t')dt' = i~\sum_{\vec{q}}\frac{|\langle 1^{\pi}_{\vk}|H_I(0)| 1^{\chi}_{\vq};1^{\pi}_{\vq+\vk}\rangle |^2}{ \omega^{\pi}(k) -\omega^{\pi}(|\vec{q}+\vk|)-\omega^{\chi}(q)+i\epsilon} \equiv~ i\,\delta E^{(2)}_\pi + \frac{\Gamma_\pi}{2} \ee   thus the imaginary part of the time integral yields the second order energy shift $\delta E^{(2)}_\pi$  and the real part yields half of the decay rate $\Gamma_\pi$ \emph{a la} Fermi's golden rule. In the case of the $\pi$ field the imaginary part vanishes by kinematics.

The matrix element is computed straightforwardly and we find
\be \mathcal{E}_{1\vec{k}} = -\frac{1}{2 \omega^{\pi}(k)}\Bigg[  \mathcal{M}^2_\pi- \frac{\lambda}{2} \int \frac{d^3q}{(2\pi^3)}\Bigg(\frac{1}{ \omega^{\pi}(q)} - \frac{1}{ \omega^{\chi}(q)}   - \frac{\lambda \sigma^2_0 }{ \omega^{\pi}(|\vec{q}+\vk|) \omega^{\chi}(q)\big(\omega^{\pi}(|\vec{q}+\vk|)+\omega^{\chi}(q) -\omega^{\pi}(k)\big)}\Bigg) \Bigg]\,. \label{enerpi}  \ee To leading order in perturbation theory one can set $\omega_\pi(k) =k$ in (\ref{enerpi}) leading to vanishing of the integral in the long wavelength limit and $\mathcal{M}_\pi=0$ from the condition (\ref{countminki}). However, keeping the $\pi$ mass selfconsistently,
in the long wavelength limit (setting $k\rightarrow 0$ in the denominator inside the integral )    the bracket   in (\ref{enerpi}) becomes
\be \Bigg[~~~\cdots ~~~\Bigg]_{k\rightarrow 0} = \mathcal{M}^2_\pi+ \frac{\lambda^2 \sigma^2_0 }{4}|\mathcal{M}_\pi| \int \frac{d^3q}{(2\pi^3)} \frac{1}{\omega^{\pi}(\vec{q}) \omega^{\chi}(q)\big(\omega^{\pi}(\vec{q})+\omega^{\chi}(q) - |\mathcal{M}_\pi| \big)} \label{masita}\ee
thus the requirement (\ref{countminki}) leads to
\be \mathcal{M}_\pi =0 \,. \label{zeromassmink}\ee

It is straightforward to check that the result (\ref{enerpi}) coincides with (\ref{massren}) for the (off-shell ) value $k=0$ in (\ref{massren}) upon integrating $q_0$ in the complex plane.

\end{document}